\documentclass[twocolumn,showpacs,superscriptaddress,twoside,pra]{revtex4}
\usepackage{mathrsfs, doi, hyperref,tabularx}
\usepackage{amssymb, amsbsy, amsmath, latexsym, dsfont, array, layout, graphicx,mathrsfs,braket,amsfonts,amsthm, amssymb,graphicx,subfigure, youngtab,color,xcolor,bm, dsfont}

\newcommand\rket[1]{\vert\, #1 )}

\hypersetup{
    colorlinks=true,       
    linkcolor=blue,          
    citecolor=blue,        
    filecolor=blue,      
    urlcolor=blue           
}

\begin{document}

\title{Coincidence Landscapes for Three-Channel Linear Optical Networks}
\author{Hubert de Guise}
\email{hubert.deguise@lakeheadu.ca}
\affiliation{Department of Physics, Lakehead University, Thunder Bay, Ontario P7B 5E1, Canada}
\author{Si-Hui Tan}
\email{sihui\_tan@sutd.edu.sg}
\affiliation{Singapore University of Technology, 20 Dover Drive, Singapore 138682}
\author{Isaac P. Poulin}
\affiliation{Department of Physics, Lakehead University, Thunder Bay, Ontario P7B 5E1, Canada}
\author{Barry C. Sanders}
\email{sandersb@ucalgary.ca}
\affiliation{
	Institute for Quantum Science and Technology, University of Calgary, Calgary, Alberta, T2N 1N4, Canada
	}
\affiliation{Program in Quantum Information Science, Canadian Institute for Advanced Research, Toronto, ON M5G 1Z8, Canada}

\begin{abstract}
We use permutation-group methods plus SU(3) group-theoretic methods
to determine the action of a three-channel passive optical interferometer on 
controllably delayed single-photon pulse inputs to each channel. 
Permutation-group techniques allow us to  relate directly expressions for rates and,
in particular,
investigate symmetries in the coincidence landscape.
These techniques extend the traditional Hong-Ou-Mandel effect analysis for two-channel interferometry
to valleys and plateaus in three-channel interferometry.
Our group-theoretic approach is 
intuitively appealing because the calculus of Wigner $D$ functions partially accounts for permutational symmetries 
and directly reveals the connections among $D$ functions, partial distinguishability, and immanants.
\end{abstract}

\date{\today}
\pacs{42.50.St,42.50.Ar,03.67.-a,03.67.Ac}
\maketitle

\section{Introduction}
\label{sec:intro}

Passive quantum optical interferometry aims to inject classical or nonclassical light into a multichannel interferometer
and count photons exiting the output ports or measure coincidences at the exits~\cite{HOM87,Cam00}
or even measure some outputs to perform post selection of the remaining output state.
This post selection procedure is a key element of the nonlinear sign gate for optical quantum computing~\cite{KLM01}
and for enhancement of the efficiency of single photons~\cite{BSSK04,BSM+04} or of single-rail optical qubits~\cite{BLS06}.
Another rapidly growing experimental direction in passive quantum optical interferometry is quantum walks~\cite{BFL+10,SCP+11,SGR+12},
which are being extended to two-photon inputs for two-walker quantum walks~\cite{SSV+12}.

The Hong-Ou-Mandel dip involves directing two identical photons such that one enters each 
of the two balanced (50:50) beam splitter inputs with  a controllable relative delay~$
\Delta$ between the pulses~\cite{HOM87}. The two photons 
can exit from the same port or different ports, and these two scenarios are distinguished 
by two-photon coincidence measurement. If both photons exit from the same port, then the 
coincidence measurement, which corresponds to the product of the measured signal at the 
two output ports, yields a 0 (essentially, two photons from one port and zero photons 
from the other yields a product $2\times 0=0$). On the other hand, one photon exiting 
each output port yields a coincidence measurement of 1 (because the yield of one photon 
from each of the two output ports results in the product $1\times 1=1$).
Measuring two-photon coincidences resulting from beam-splitter mixing of two single 
photons underpins much of the field of passive quantum optical interferometry.
The term passive is used to distinguish quantum optical interferometry from incorporation of active elements within the interferometer such as linear or parametric amplifiers.

The Hong-Ou-Mandel dip is a decrease in the two-photon coincidence rate~$\wp$ near zero 
delay between identical single photons at each port of a balanced (50:50) beam splitter.
This dip can be generalized to more than two channels and to an extension from injecting 
single photons into each input port to the case of single photons entering some
ports and nothing (vacuum) entering other input ports.

Higher-order coincidence dips could be observed by placing detectors at several output 
ports. Suppose output ports $i$, $j$, $k$, and $l$ each lead to a photo detector
and four photons are injected into the interferometer. Then the detectors can ``see'' 
$m_i$, $m_j$, $m_k$, and $m_l$ photons, respectively, such that $\sum_{a\in\{i,j,k.l\}} 
m_a\leq 4$, where the inequality is saturated only if the photons do not exit
other ports or are lost by the detectors.

The four-photon coincidence product is then 
$$\prod_{a\in\{i,j,k.l\}} m_a,$$
which is 0 except for the case where exactly one photon 
leaves each of the output ports. This product of counts from specified ports, such that a 
nonzero value is only obtained if a single photon exits each port, is the generalization 
of the Hong-Ou-Mandel dip two-photon coincidence rate for multiple channels, several 
single-photon inputs, and multiphoton coincidence detection.

Recently generalizing the Hong-Ou-Mandel dip has been the subject of considerable interest because of the BosonSampling problem. The BosonSampling problem demands sampling of the output photon coincidence distribution given an interferometric input comprising single-photon and vacuum states.
The output coincidence distribution is computationally hard to sample classically but efficiently simulatable with a quantum
optical interferometer (subject to some conjectures and an assumption about scalability)~\cite{AA11}.
The BosonSampling problem has led to several reports of experimental success
based on generalizing the Hong-Ou-Mandel dip~\cite{MTS+12,BFR+13,SMH+13,TDH+13,COR+13,RMP+}
(including experimental verification~\cite{SVB13,CS13}).

Theoretical analysis of the generalized Hong-Ou-Mandel dip typically focuses on 
simultaneous arrival of the identical photons. With arbitrary delays between photons, the 
Hilbert space~$\mathscr{H}$ for the system is large because single-mode treatments of 
input photons give way to an infinite (temporal) mode continuum for each input photon.
Another complication in studies of generalized Hong-Ou-Mandel dips is that the number~$m$ 
of channels can exceed the number~$n$ of photons ($m\ge n$). In this general case 
the Hilbert space dimension is $\text{dim}\mathscr{H}=m^n$. On the other hand, when all 
delays between photons are $0$ so each photons can be treated within the single-mode 
framework and the set of all photons is symmetric under exchange, the only subspace of 
the full Hilbert space with nonzero support (i.e., the largest subspace such that the 
overlap of states in this subspace with the multi-photon state is not zero) is the 
subspace of states fully symmetric under permutation of frequencies; the  Hilbert spaces dimension is then exponentially smaller:
$${m+n-1\choose n}.$$

This emphasis on simultaneity for higher-order Hong-Ou-Mandel dip contrasts sharply with 
experimental practice for the standard Hong-Ou-Mandel dip, which utilizes a controllable 
time delay~$\tau$ between the two photons. Controlling~$\tau$ is essential to verify that 
the dip is behaving approximately as expected and, furthermore, to calibrate the extent 
of the dip relative to the background coincidence rate.
 Ideas in this direction 
have also been developed for two photons arriving
in each of the beam-splitter ports~\cite{RTLKMBK2012}.

Some of us recently showed that nonsimultaneity breaks the full permutation symmetry of the 
input state~\cite{TGdGS13}. This broken permutation symmetry causes the output 
coincidence rate to depend on immanants~\cite{LR34,Lit50,Val79,Bue00} of the 
interferometer transition matrix. The immanant is a generalization of the permanent, 
which is relevant for permutation-symmetric input states, and the determinant, which 
holds for the antisymmetric case.

Our previous work focused on determining and explaining the ``coincidence landscape'' for three-channel passive optical
interferometry with single photons injected into each of three input ports.
Each photon can be delayed independently and controllably.
The time delay vector
\begin{equation}
\label{eq:tau}
	\bm{\tau}:=(\tau_1,\tau_2,\tau_3)
\end{equation}
represents the time delays for the first, second, and third photon, respectively.
An overall time reference frame can be ignored so only two time delays are required,
given by the two-component relative vector
\begin{equation}
\label{eq:Delta}
	\bm{\Delta}:=(\Delta_1,\Delta_2)=(\tau_2-\tau_1,\tau_3-\tau_2).
\end{equation}
Therefore,
\begin{equation}
\label{eq:tauDelta}
	\bm{\tau}=(\tau_1,\tau_1+\Delta_1,\tau_{2}+\Delta_2),
\end{equation}
The coincidence rate~$\wp(\bm{\Delta})$ is thus a function of two independent variables 
and can be represented as a surface plot. This landscape is observed experimentally, 
albeit in more complicated three-photon-through-five-channel interferometry suitable for 
first-principle tests of experimental BosonSampling~\cite{MTS+12,BFR+13,SMH+13,TDH+13,COR+13,SVB13,RMP+}.

Our earlier brief results in three-channel interferometry with single photons injected 
into each input port indicate the role of immanants but do not delve into rich aspects of 
the coincidence landscapes. Our aim here is to present the following new results as well 
as to clarify subtleties in the earlier work. We provide a thorough, comprehensive 
explanation of the three-photon coincidence rate~$\wp$ with controlled timing delays~
$\bm\Delta$. Earlier we determined coincidence landscapes based on photon counting 
operators that were dual to the source-field operators~\cite{TGdGS13};
this time we forgo the mathematically elegant dual approach in favor of the detector 
model matching current experimental implementations~\cite{RR06}. Also we analyze and 
explain the extremal cases for which two photons arrive simultaneously and one arrives 
significantly later or earlier so is distinguishable. Furthermore, we study the case where 
all three photons are distinct due to long pairwise time delays.

Our analysis serves to explain in detail the three-photon generalized Hong-Ou-Mandel dips and its extreme cases of complete distinguishability of one or all photons.
This work not only lays a foundation for generalizing the Hong-Ou-Mandel dip theory beyond the three-photon level
but also emphasizes the connection between these generalized dips and immanants,
thus extending the paradigm of the BosonSampling problem from permanents to immanants.
Our group-theoretic methods elucidate the role of immanants in the features of the photon-coincidence landscape beyond the simultaneous--photon-arrival limit and furthermore, exploits SU(3) group-theoretic properties in the photon-number-conserving case to reduce the overhead for calculating and numerically computing photon-coincidence rates compared to not using these relations.
The reduction in computational cost
(but not a reduction in algorithmic complexity scaling per se)
by using our methods, instead of brute-force techniques
based on working directly with mappings of creation operators according to the interferometer transition matrix,
is due in part to the built-in exploitation of permutation symmetries in our mathematical framework.

\section{Two input ports and SU(2)}
\label{sec:from2photons}

Although the focus of this work is on three-channel passive quantum optical interferometry with one photon
injected into each input port,
a thorough understanding of the humble beam splitter (balanced or otherwise) is needed first.
The reason for this need is that the beam splitter is the basic building block of general passive quantum optical interferometric transformations.
Despite its simplicity, the beam splitter still holds surprises such as the recent universality proof for beam splitters~\cite{BA14}.

\subsection{Two monochromatic photons} 
\label{sec:2monochromatic}

In this section
we expound on the example of two monochromatic photons
entering a four-port system, i.e., two input ports and two output ports.
For~$\hat{a}_{j,\text{in}}^\dagger(\omega)$
the creation operator for an input monochromatic photon in mode~$j$,
a monochromatic single-photon state in mode~$j$ is
\begin{equation}\label{monoph}
	\rket{1(\omega)_j}:=\hat{a}_j^\dag(\omega)\ket{0},\, j=1,2.
\end{equation}
We use the parenthetical (rounded) bra-ket notation to denote the purely monochromatic states.
The commutator relation for monochromatic creation and annihilation operators is
\begin{equation}
\label{eq:commutator}
	[\hat{a}_k(\omega_i),\hat{a}^\dag_\ell(\omega_j)]=\delta_{k\ell}\delta(\omega_i-\omega_j)\mathds{1}\ ,
\end{equation}
with~$\mathds{1}$ the identity operator.

A beam splitter is equivalent to a four-port passive interferometer,
with two input ports and two output ports.
Mapping the two input modes to the two output modes is achieved by the photon-number-conserving 
linear transformations scattering single-input to single-output photons as
\begin{align}
	\hat a_{1,\text{out}}^\dagger(\omega)&= U_{11}\hat{a}_{1,\text{in}}^\dagger(\omega)+
	U_{12}\hat{a}_{2,\text{in}}^\dagger(\omega)\nonumber , \\
	 \hat a_{2,\text{out}}^\dagger(\omega)&= U_{21}\hat{a}_{1,\text{in}}^\dagger(\omega)+
	U_{22}\hat{a}_{2,\text{in}}^\dagger(\omega),
\end{align}
from which
\begin{equation}
	\bm{U}=	\begin{pmatrix}
				U_{11}&U_{12}\\U_{21}&U_{22}
			\end{pmatrix}
\end{equation}
is constructed
with each entry~$U_{ij}$ in the matrix $\bm{U}$ being treated as a frequency-independent quantity.
In practical optical systems,
this assumption is desirable and approximately valid for narrow-band optical fields.

Conservation of total photon number of photons requires that
\begin{equation}
	\bm{U}^\dagger \bm{U}=\bm{U} \bm{U}^\dagger=\mathds{1};
\end{equation}
hence $\bm{U}$ is unitary with determinant det$\bm{U}=\text{e}^{\text{i}\varphi}$.
Therefore,
$\bm{U}$ can be rewritten as
\begin{equation}
\bm{U}=	R(\Omega) \cdot
	\begin{pmatrix}
		\text{e}^{\text{i}\varphi/2}&0\\
		0&\text{e}^{\text{i}\varphi/2}
	\end{pmatrix}, \label{factorU}
\end{equation}
with the matrix
\begin{align}
R(\Omega)
&=
\begin{pmatrix}
	\text{e}^{-\frac{1}{2}\text{i}(\alpha +\gamma )} \cos \frac{\beta}{2} & -\text{e}^{-\frac{1}{2}\text{i}(\alpha -\gamma )} \sin \frac{\beta}{2} \\
	\text{e}^{\frac{1}{2}\text{i}(\alpha -\gamma )} \sin \frac{\beta}{2} & \text{e}^{\frac{1}{2}\text{i}(\alpha +\gamma )} \cos \frac{\beta}{2} \\
\end{pmatrix}
\label{eq:2x2Rmatrix}
\end{align}
special and unitary, i.e., unitary with determinant $+1$ depending on the three parameters
\begin{equation}
\label{eq:Omega}
	\Omega:=(\alpha,\beta,\gamma).
\end{equation}

The matrix~$R$ in Eq.~(\ref{eq:2x2Rmatrix}) is a three-parameter $2\times 2$ special 
unitary matrix. The unitarity of the matrix is evident by computing $RR^\dagger$ and 
$R^\dagger R$ and obtaining the $2\times 2$ identity matrix~$\mathds{1}$ in each case.
Therefore, $R\in$U($2$). Furthermore, det$R=1$ so~$R$ is ``special'', implying that $R\in
$SU($2$). The fact that the group SU(2) represents the action of the beam splitter is 
because the passive lossless beam-splitter preserves the total photon number: 
the number of photons entering the beam-splitter input ports equals the number of photons 
exiting the beam-splitter output ports. The $2\times 2$ matrix representation arises for 
just one photon entering the beam-splitter. In general, the matrices are of size 
$(2j+1)\times (2j+1)$ for a total of~$2j$ photons entering the beam-splitter, with $j$ 
either an integer or a half-odd integer~\cite{CST89}.

We now introduce the SU(2) $D$ function for the irreducible representation (irrep)~$j$,
\begin{equation}
	D^j_{mm'}(\Omega)
		:=\langle jm|\text{e}^{-i\alpha \hat J_z}\text{e}^{-i\beta \hat J_y}\text{e}^{-i\gamma \hat J_z}|jm'\rangle , \label{defineD}
\end{equation}
with $\hat J_k$ the $(2j+1)\times (2j+1)$ matrix representation of the three 
$\mathfrak{su}(2)$ algebra generators. 
We use a standard abuse of language and refer to the $\mathfrak{su}(2)$ algebra generators as 
``angular momentum operators'', although there is no connection with the physical angular momentum.
In other words, the analogy with angular momentum reflects the abstract notation that the 
beam-splitter action on incoming photons can be regarded as an abstract photon-number-
preserving ``rotation'' of the photonic state in the two (input or output) modes.

From this formalism of ``angular momentum'' operators $\{\hat J_k\}$,
we see that~$\Omega$ (\ref{eq:Omega}) is just the Euler-angle triplet for the SU(2) transformation
$$\text{e}^{-i\alpha \hat J_z}\text{e}^{-i\beta \hat J_y}\text{e}^{-i\gamma \hat J_z},$$
and the entries 
of the matrix $R(\Omega)$ are Wigner $D$ functions for the SU(2) representation of $j=1/2$:
\begin{equation}
\label{eq:basicsu2matrix}
	R(\Omega)
		=\begin{pmatrix}
			D^{\frac{1}{2}}_{\frac{1}{2},\frac{1}{2}}(\Omega)&D^{\frac{1}{2}}_{\frac{1}{2},-\frac{1}{2}}(\Omega)\\
			D^{\frac{1}{2}}_{-\frac{1}{2},\frac{1}{2}}(\Omega)&D^{\frac{1}{2}}_{-\frac{1}{2},-\frac{1}{2}}(\Omega)
		\end{pmatrix}.
\end{equation}
General  expressions for $D^j_{mm'}(\Omega)$ are known,
and tables of explicit functions for various~$j$ and~$m$ and~$m'$ exist~\cite{VMK88}.

If two photons of frequencies $\omega_a$ and $\omega_b$ 
enter ports~$1$ and $2$, respectively, and exit in distinct ports 
-- say ports  $2$ and~$1$ -- this post selected output state is constructed by applying the product
\begin{equation}
	\hat{a}_{1,\text{out}}^\dagger(\omega_b)\hat{a}_{2,\text{out}}^\dagger(\omega_a)
		=\left[U_{21}\hat{a}_{1,\text{in}}^\dagger(\omega_a)\right]\left[U_{12}\hat{a}_{2,\text{in}}^\dagger(\omega_b)\right]
\label{1b2a}
\end{equation}
to the vacuum.
The diagonal matrix on the right-hand side of Eq.~(\ref{factorU}) is constant so, for this specific case,
\begin{align}
\label{scattering2photons}
	&\hat{a}_{1,\text{out}}^\dagger(\omega_b)\hat{a}_{2,\text{out}}^\dagger(\omega_a)\nonumber \\
		&\  =\text{e}^{\text{i}\varphi}D^{\frac{1}{2}}_{-\frac{1}{2},\frac{1}{2}}(\Omega)D^{\frac{1}{2}}_{\frac{1}{2},-\frac{1}{2}}(\Omega)
		\hat{a}_{1,\text{in}}^\dagger(\omega_a)\hat{a}_{2,\text{in}}^\dagger(\omega_b).
\end{align} 
The overall phase $\text{e}^{\text{i}\varphi}$ is not of operational importance and, hence, safely discarded.

We employ the obvious relationship
\begin{align}
\label{eq:u21u12}
	D^{\frac{1}{2}}_{-\frac{1}{2},\frac{1}{2}}D^{\frac{1}{2}}_{\frac{1}{2},-\frac{1}{2}}
		=&\frac{1}{2}\left(D^{\frac{1}{2}}_{-\frac{1}{2},\frac{1}{2}}D^{\frac{1}{2}}_{\frac{1}{2},-\frac{1}{2}}-
		D^{\frac{1}{2}}_{\frac{1}{2},\frac{1}{2}}D^{\frac{1}{2}}_{-\frac{1}{2},-\frac{1}{2}}\right)\nonumber\\
		&+ \frac{1}{2}\left(D^{\frac{1}{2}}_{-\frac{1}{2},\frac{1}{2}}D^{\frac{1}{2}}_{\frac{1}{2},-\frac{1}{2}}+
		D^{\frac{1}{2}}_{\frac{1}{2},\frac{1}{2}}D^{\frac{1}{2}}_{-\frac{1}{2},-\frac{1}{2}}\right)
\end{align}
where the explicit dependence of each term on~$\Omega$ is suppressed.
Henceforth, we suppress explicit $\Omega$ dependence when the nature of the dependence on~$\Omega$ is self-evident.
The first term on the right-hand side of Eq.~(\ref{eq:u21u12}) is
\begin{equation}
\label{eq:detU}
	D^{\frac{1}{2}}_{-\frac{1}{2},\frac{1}{2}}D^{\frac{1}{2}}_{\frac{1}{2},-\frac{1}{2}}-
		D^{\frac{1}{2}}_{\frac{1}{2},\frac{1}{2}}D^{\frac{1}{2}}_{-\frac{1}{2},-\frac{1}{2}}
		=-\text{det}\,R=-1
\end{equation}
for $R(\Omega)$ given in Eq.~(\ref{eq:basicsu2matrix}).

For the second term on the right-hand side of Eq.~(\ref{eq:u21u12}),
we resort to the formula for the permanent of a matrix $2\times 2$ matrix $X:(x_{ij})$,
which is
\begin{equation}
\label{eq:2x2per}
	\text{per}X=x_{11}x_{22}+x_{12}x_{21}.
\end{equation}
Thus,
\begin{align}
\label{eq:perU}
&	D^{\frac{1}{2}}_{-\frac{1}{2},\frac{1}{2}}D^{\frac{1}{2}}_{\frac{1}{2},-\frac{1}{2}}+
		D^{\frac{1}{2}}_{\frac{1}{2},\frac{1}{2}}D^{\frac{1}{2}}_{-\frac{1}{2},-\frac{1}{2}}\nonumber \\
		&=\cos\beta=\text{per}R.
\end{align}

From the expressions for the determinant and permanent of a $2\times 2$ matrix,
we can rewrite the amplitude in  Eq.~(\ref{scattering2photons}) as
\begin{equation}
\label{eq:amplitude}
	\text{e}^{\text{i}\varphi}D^{\frac{1}{2}}_{-\frac{1}{2},\frac{1}{2}}(\Omega)D^{\frac{1}{2}}_{\frac{1}{2},-\frac{1}{2}}(\Omega)
\end{equation}
so  the scattering
\begin{equation}
\label{eq:scattering}
	\hat{a}_{1,\text{in}}^\dagger(\omega_a)\hat{a}_{2,\text{in}}^\dagger(\omega_b)
		\to\hat{a}_{1,\text{out}}^\dagger(\omega_b)\hat{a}_{2,\text{out}}^\dagger(\omega_a)
\end{equation}
can be written, up to an overall and unimportant $\text{e}^{\text{i}\varphi}$, as
\begin{equation}
	D^{\frac{1}{2}}_{-\frac{1}{2},\frac{1}{2}}D^{\frac{1}{2}}_{\frac{1}{2},-\frac{1}{2}}
		=\frac{1}{2}\left(\text{per}\,R-\text{det}\,R\right).
\label{su21su12}
\end{equation}
Equation~(\ref{su21su12}) shows, on general grounds, that 
the amplitude for coincidence counts in distinct output ports can be written in terms
of the permanent and the determinant of the matrix $R(\Omega)$;  these are in turn expressed as combinations of
products of elements of the matrix $R(\Omega)$.

Consider, instead of Eq.~(\ref{1b2a}), scattering of the input state 
$\hat{a}_{1,\text{in}}^\dagger(\omega_a)\hat{a}_{2,\text{in}}^\dagger(\omega_b)\ket{0}$ to the other possible post selected output state
\begin{align}
	&\hat{a}_{1,\text{out}}^\dagger(\omega_a)\hat{a}_{2,\text{out}}^\dagger(\omega_b)\ket{0}
			\nonumber \\&
		=\left[U_{11}\hat{a}_{1,\text{in}}^\dagger(\omega_a)\right]\left[U_{22}\hat{a}_{2,\text{in}}^\dagger(\omega_b)\right]\ket{0}
\end{align}
with the resulting scattering amplitude
\begin{equation}
\text{e}^{\text{i}\varphi}D^{\frac{1}{2}}_{\frac{1}{2},\frac{1}{2}}(\Omega)
		D^{\frac{1}{2}}_{-\frac{1}{2},-\frac{1}{2}}(\Omega)=\frac{\text{e}^{\text{i}\varphi}}{2}\left(\text{per}\,R+\text{det}\,R\right).
		\label{1a2b}
\end{equation}

This scattering amplitude is related to that resulting from Eq.~(\ref{1b2a}) as follows.
The output states are related by a permutation of the frequencies of photons in modes 1 and 2.  
To this end, we introduce the matrix
\begin{equation}
\label{eq:R21matrix}
	R^{21}
		=\begin{pmatrix}
		D^{\frac{1}{2}}_{-\frac{1}{2},\frac{1}{2}}&D^{\frac{1}{2}}_{-\frac{1}{2},-\frac{1}{2}}\\
		D^{\frac{1}{2}}_{\frac{1}{2},\frac{1}{2}}&D^{\frac{1}{2}}_{\frac{1}{2},-\frac{1}{2}}
\end{pmatrix}, 
\end{equation}
which is obtained by permuting rows of $R(\Omega)$ in Eq.~(\ref{eq:basicsu2matrix}) so that row 1 of $R(\Omega)$ 
is row 2 of $R^{21}(\Omega)$,
and row 2 of $R(\Omega)$ is row 1 of $R^{21}(\Omega)$.
In fact, we can rewrite Eq.~(\ref{su21su12})  as
\begin{equation}
	D^{\frac{1}{2}}_{-\frac{1}{2},\frac{1}{2}}D^{\frac{1}{2}}_{\frac{1}{2},-\frac{1}{2}}
		=\frac{1}{2}\left(\text{per}\,R^{21}+\text{det}\,R^{21}\right).
\label{su2scattering21}
\end{equation}
We see that this scattering amplitude has the same form (up to an overall phase) 
as the amplitude of Eq.~(\ref{1a2b}) (they are ``covariant'').
The scattering amplitude of Eq.~(\ref{su2scattering21}) is obtained (up to an overall phase) by substituting~$R$ with~$R^{21}$ in  Eq.~(\ref{1a2b}).
Thus, permuting the output frequencies induces a permutation of the rows that transforms $R\to R^{21}$
but does not change the expression of the scattering amplitude when written in terms of the permanent and the
determinant.
 
Another observation is linked to the permutation of frequencies.
The effect of such a permutation
can also be made explicit by introducing the states
\begin{equation}
\label{eq:Psi+-}
	| \Psi^{\pm}\rangle
		=\frac{1}{\sqrt{2}}\left(\hat a^\dagger_{1}(\omega_a)\hat a^\dagger_2(\omega_b)
			\pm\hat a^\dagger_{1}(\omega_b)\hat a^\dagger_2(\omega_a)\right)
\vert 0\rangle, 
\end{equation}
which are clearly symmetric and antisymmetric with respect to
the permutation group~$S_2$ for the two frequencies~$\omega_{a,b}$.
States~(\ref{eq:Psi+-}) are, respectively,
the $\ell=1,m=0$ (triplet) state and the $\ell=0,m=0$ (singlet) state,
which can be obtained from the usual theory of two-mode systems in terms of angular momentum.
The interferometric input and output states can be expanded in terms of  $\vert \Psi^\pm\rangle $,
and the effect of permuting frequencies of the output state is determined from the permutation of frequencies on $\vert \Psi^\pm\rangle $.

The group $S_2$ contains two elements
represented by the identity~$\mathds{1}$ and the permutation~$P_{12}$, which exchanges 
$\omega_1$ with~$\omega_2$). 
Representations of $S_2$ are conveniently labeled using the method of Young 
diagrams~\cite{Wyb70,Tun85,FH91,Yon07}:  
$\Yboxdim{6pt}\yng(2)$ for the symmetric representation, 
and $\Yboxdim{6pt}\yng(1,1)$ for the antisymmetric representation.

We emphasize the role of the permutation group by writing
\begin{equation}
	\left|\Psi^+\right\rangle\to \left|\Psi^{\Yboxdim{4pt}\yng(2)}\right\rangle,
	\left| \Psi^-\right\rangle\to \left|\Psi^{\Yboxdim{4pt}\yng(1,1)}\right\rangle;
\end{equation}
i.e., we can explicitly identify the $\ell=1$ SU(2) state $\vert \Psi^+\rangle$ with one of the basis states for the 
$\Yboxdim{6pt}\yng(2)$ representation of $S_2$, and the $\ell=0$ SU(2) state $\vert \Psi^-\rangle$ 
with the basis state for the $\Yboxdim{6pt}\yng(1,1)$ representation.  This identification of representations of the
symmetric group and representations of SU(2) is an example of Schur-Weyl duality~\cite{Wey39,Lic78,RW10},
which proves invaluable in our discussion of SU(3) irreps later.

The scattering amplitudes 
\begin{align}
	\langle \Psi^{\Yboxdim{4pt}\yng(2)}\vert R \vert\Psi^{\Yboxdim{4pt}\yng(2)}\rangle
		=&D^{\frac{1}{2}}_{\frac{1}{2},\frac{1}{2}}D^{\frac{1}{2}}_{-\frac{1}{2},-\frac{1}{2}}
			+D^{\frac{1}{2}}_{-\frac{1}{2},\frac{1}{2}}D^{\frac{1}{2}}_{\frac{1}{2},-\frac{1}{2}},\nonumber \\
		=&\cos\beta
		=\text{per}R =D^{1}_{00}\equiv D^{\Yboxdim{4pt}\yng(2)}_{00},  \label{D1asasum}
\end{align}
and
\begin{align}
	\langle \Psi^{\Yboxdim{4pt}\yng(1,1)}\vert R \vert \Psi^{\Yboxdim{4pt}\yng(1,1)}\rangle
		=&
D^{\frac{1}{2}}_{\frac{1}{2},\frac{1}{2}}D^{\frac{1}{2}}_{-\frac{1}{2},-\frac{1}{2}} 
{ -}D^{\frac{1}{2}}_{-\frac{1}{2},\frac{1}{2}}D^{\frac{1}{2}}_{\frac{1}{2},-\frac{1}{2}}\nonumber \\
		=&1
		=\text{det}R=D^{0}_{00}\equiv D^{\Yboxdim{4pt}\yng(1,1)}_{00}
\label{D0asasum} 
\end{align}
are easily verified. In the case of the 50-50 beam-splitter, $\beta=\pi/2$ such that the scattering amplitude for indistinguishable single photons in Eq.~(\ref{D1asasum}) vanishes.
Thus, we have determined the roles of the permanent and determinant of the matrix $R$, the $S_2$ permutation symmetry of the input state, and
higher $D$ functions of the group SU(2) for the two-channel interferometer and the beam-splitter.
Higher $D$ functions arise as linear combinations of products of basic $D^{1/2}$ functions entering
in the $2\times 2$ matrix $R$ of Eq.~(\ref{eq:2x2Rmatrix}), as per Eqs.~(\ref{eq:detU}) and (\ref{eq:perU}).

Finally, we can use Young diagrams to summarize neatly the contents of Eqs.~(\ref{1a2b}) and (\ref{su2scattering21}) so as to display
the covariance explicitly:
\begin{equation}
	\bra{0}a_{i}(\omega_a)a_{j}(\omega_b)R\,a^\dagger_{1}(\omega_a)a^\dagger_{2}(\omega_b)\ket{0}
		=\textstyle\frac{1}{2}{\Yboxdim{6pt}\yng(2)}_{ij}+\textstyle\frac{1}{2}{\Yboxdim{6pt}\yng(1,1)}_{ij},
\label{su2covariantrate}
\end{equation}
where ${\Yboxdim{4pt}\yng(2)}_{ij}:=\text{per}\,R^{ij}$ is the permanent of the matrix $R^{ij}$, and where
 ${\Yboxdim{4pt}\yng(1,1)}_{ij}:=\text{det}\,R^{ij}$ is the determinant of $R^{ij}$.

\subsection{Pulsed sources and finite-bandwidth detectors}
\label{subsec:pulsed}

For realistic systems the input photons are not monochromatic, nor should they be.
If photons are to be delayed relatively to each other,
their temporal envelopes need to be of finite duration.
This temporal-mode envelope is sometimes called the photon wavepacket.
Detectors are not strictly monochromatic either, as the duration of the detection must be finite.
In practice, detectors are often preceded by spectral filters so are close to monochromatic.
Mathematically the source field and the detector response should be modeled not as monochromatic functions as in the previous section
but, rather, in terms of the appropriate temporal mode.

Mathematically the state of one photon in each of~$n$ modes is a complex-value-weighted multifrequency integral
of monochromatic single-photon states in each mode~(\ref{monoph}) given by
\begin{equation}
\label{eq:n}
	\ket{n}
		=\frac{1}{\sqrt{n!}}\int\text{d}^n\bm{\omega}
		\tilde{\phi}(\omega_1)\cdots \tilde{\phi}(\omega_n) |1(\omega_1)_1\ldots 1(\omega_n)_n) \ ,
\end{equation}
for $\bm{\omega}:=(\omega_1,\dots,\omega_n)$ the $n$-dimensional frequency,
d$^n\bm{\omega}$ the $n$-dimensional measure over this domain,
and~$\tilde{\phi}(\omega)$ the spectral function.
For the source spectral function,
we choose a Gaussian~\cite{TGdGS13}
\begin{equation}
\label{eq:phitilde}
	\tilde{\phi}(\omega)
		=\frac{1}{\left(2\pi\sigma_0^2\right)^{1/4}}
			\exp\left[-\frac{(\omega-\omega_0)^2}{4\sigma_0^2}\right]
\end{equation}
with~$\omega_0$ the carrier frequency and~$\sigma_0$ the half-width of the Gaussian distribution.

Previously we treated the detector as dual~\cite{TGdGS13} in the sense that photon counting corresponds to the Fock number-state projector~$|n\rangle\langle n|$,
for~$|n\rangle$ the number state~(\ref{eq:n}),
but here we employ the flat-spectrum incoherent Fock-number state measurement operator~\cite{RR06}
\begin{align}
\label{eq:Mpovm}
	\int\text{d}^n\bm{\omega}|1(\omega_1)_1\ldots 1(\omega_n)_n )(1(\omega_1)_1\cdots 1(\omega_n)_n|
\end{align}
that is applicable to detectors currently used in photon-coincidence experiments
such as the BosonSampling type~\cite{MTS+12,BFR+13,SMH+13,TDH+13,COR+13,RMP+}.
A further adjustment accounts for the threshold nature of single-photon counting modules:
due to saturation they either see nothing or measure inefficiently at least one photon without number-resolving capability~\cite{BS02}.

In the case of the two-photon Hong-Ou-Mandel dip experiment~\cite{HOM87},
the coincidence rate is a linear combination of the determinant,~(\ref{eq:detU}), and permanent,~(\ref{eq:perU}), of the 
$2\times 2$ matrix.  The coefficients of this combination are 
controlled through an adjustable time delay~$\tau$
between the pulses arriving at respective times~$\tau_1$ and~$\tau_2$ such that $\Delta:=\tau_1-\tau_2$.
For identical Gaussian pulses of unit width ($\sigma_0=1$) 
and the detector measurement,~(\ref{eq:Mpovm}),
the resultant coincidence rate is
\begin{align}
\label{eq:P11}
	\wp_{11}(\Delta)
		=&\frac{1}{2}\Bigg[\left(1+\text{e}^{-\Delta^2}\right)\left|D^{\Yboxdim{4pt}\yng(2)}_{00}(\Omega)\right|^2\nonumber\\
			&+\left(1-\text{e}^{-\Delta^2}\right)\left|D^{\Yboxdim{4pt}\yng(1,1)}_{00}(\Omega)\right|^2\Bigg].
\end{align}

For zero time delay $\Delta=0$,
the pulses are indistinguishable.
For~$\Delta=0$,
Eq.~(\ref{eq:P11}) reduces to
\begin{equation}
\label{eq:P11reduced}
	\wp_{11}(\Delta)
		=\left|D^{\Yboxdim{4pt}\yng(2)}_{00}(\Omega)\right|^2.
\end{equation}
Thus, the antisymmetric part~$D^{\Yboxdim{4pt}\yng(1,1)}_{00}(\Omega)$ of Eq.~(\ref{eq:P11}) has vanished,
and only the symmetric part of the amplitude $D^{\Yboxdim{4pt}\yng(2)}_{00}(\Omega)$ survives:
$D^{\Yboxdim{4pt}\yng(2)}_{00}(\Omega)=\cos\beta$.
The balanced beam splitter has $\beta=\pi/2$
so $D^{\Yboxdim{4pt}\yng(2)}_{00}(\Omega)=0$ thereby leading to $\wp_{11}(\Delta=0)=0$.

This null amplitude thus results in the Hong-Ou-Mandel dip corresponding 
to a nil coincidence rate at $\Delta=0$ in the ideal limit; i.e., the nil 
coincidence rate shows that the two photons entering the beam splitter are 
operationally indistinguishable. Under these conditions the two photons 
are forbidden to yield a coincidence because the probability amplitudes 
for the two cases where both photons are transmitted and both photons are 
reflected (the two cases that would yield a coincidence) cancel each other, 
leaving only the noncoincidence events of both photons being in a 
superposition of going one way or the other.

\section{Three monochromatic photons and SU(3)}
\label{sec:SU3}

In this section we establish the necessary notation and develop the mathematical framework concerning three monochromatic photons,
each entering a different input port of a passive three-channel optical interferometer
and undergoing coincidence detection at the three output ports.
As in the previous section,
arguments such as~$\Omega$ for transformations $R$ and functions $D$ are suppressed when obvious so as not
to overcomplicate the expressions and equations.

\subsection{Preface}
\label{subsec:preface}

In Sec.~\ref{subsec:interferometrictransformation} we generalize 
Eq.~(\ref{factorU}) to the case $3\times 3$ matrices.
The SU(2) $R$ matrix of Eq.~(\ref{factorU}) become an SU(3) matrix.  
We report in Appendix \ref{appendixsu3} essential details on the Lie algebra $\mathfrak su$(3) and
their representations~\cite{BB63,Sla81}.
Representations of SU(3) are obtained by exponentiating the corresponding representations of $\mathfrak{su}$(3).
In Sec.~\ref{subsec:Dreps} we briefly discuss the SU(3) $D$ functions using standard
labeling and construction for SU(3) $D$ functions~\cite{RSdG99}.

We employ appropriate basis states,
endowed with ``nice'' properties under permutation of output modes,
to obtain the $D$ functions~\cite{RSdG99}.
Some of these ``nice'' properties are given explicitly in 
Eqs.~(\ref{I230state}) and (\ref{I231state}) in Appendix \ref{appendixsu3}.
The required states are either symmetric or antisymmetric under permutation of modes~2 and~3.
The action of elements of the permutation group of three objects ($S_3$) on these basis states and thus on the 
$D$ functions has been discussed earlier~\cite{RSdG99}.

The connection between scattering
amplitudes and $D$ functions is given in Eq.~(\ref{eq:productofSU3Ddecomposed}) and Table~\ref{table:tableofcijk}.
As in Sec.~\ref{sec:from2photons}, we eventually label the SU(3) irreps with Young diagrams.
For an interferometer containing 3 photons, the Young diagram have 3 boxes.  Young diagrams with 3 boxes also label representations of the permutation
group $S_3$.

We briefly discussed in Subsec.~\ref{sec:2monochromatic} the effect on scattering amplitudes of permuting two of the output photons.
An important part of our work is to generalize this discussion to the three-photon case.
We start this in Subsec.~\ref{subsec:S3partitionsimmanants} where we introduce the permutation group $S_3$ of three objects.

The permutation group $S_3$ has a richer structure than does the permutation group of two objects.
In addition to defining the permanent and the determinant of a $3\times 3$ matrix,
we define additionally another type of matrix function known as an immanant~\cite{LR34,Lit50,Val79,Bue00}.
The permanent and the determinant are in fact special cases of immanants.
The immanants of the $3\times 3$ matrix $R$ are 
constructed as linear combinations containing in general six triple products of entries of $R$.
Here we provide few explicit expression as the expressions are excessively complicated to include in full.
 Whereas the permanent and determinant can be expressed in terms of a single SU(3) $D$-function,
thereby generalizing Eqs.~(\ref{D1asasum}) and~(\ref{D0asasum}) of the previous section,
the last immanant of the SU(3) matrix is a linear combination of SU(3) $D$-functions 
as introduced earlier~\cite{RSdG99}.

In Subsec.~\ref{subsec:Dfunctionsimmanants}
we generalize our previous discussion of the effect of permutation of frequencies on rates in the two-mode problem
to the effect of permuting the frequencies for the three-mode case.
We also discuss the connection between $D$-functions and immanants of matrices $R^{ijk}$ where rows have 
been permuted.
The permanent and determinant transform back into themselves (up to maybe a sign in the case of the determinant) under 
such a permutation of rows.
In general immanants do not satisfy such a simple relation:
their transformation rules are more complicated.  
We provide in Eq.~(\ref{eq:immD}) of Subsec.~\ref{subsec:Dfunctionsimmanants}
the explicit expression of the ${\Yboxdim{6pt}\Yvcentermath1 \yng(2,1)}_{ijk}$ immanants in terms of SU(3) 
$D$-functions~\cite{TGdGS13}.

Finally, in Sec.~\ref{subsec:amplitudesimmanants}, we provide details on the relation between rates and immanants.
Just as the scattering rate for two monochromatic photons can be expressed in terms of the permanent and 
the determinant of 
the appropriate $2\times 2$ scattering matrix, the scattering rate for three monochromatic 
photons can be written
in terms of the immanants of the appropriate $3\times 3$ scattering matrix.   

We have shown explicitly in Sec.~\ref{sec:2monochromatic} how the scattering rates can 
be written in a covariant form by using the permanent and determinant of the $2\times 2$ matrix $R(\Omega)$.
Previously we found in~\cite{TGdGS13} that the same observation holds for the case of three photons in a three-channel interferometer.  
This result is summarized in  Eq.~(\ref{eq:weightedsum}), which leads to the result that, for monochromatic photons, 
the rates have simple expressions in terms of immanants of the matrices~$R^{ijk}$.

\subsection{The interferometric transformation}
\label{subsec:interferometrictransformation}

A general interferometer with three input and three output ports transforms the creation operators for input photons to the output creation operators, where the action on the basis vectors of each photon is
\begin{equation}
	\begin{pmatrix}
		\hat{a}_{1,\text{out}}^\dagger(\omega)\\ \hat{a}_{2,\text{out}}^\dagger(\omega) \\ \hat{a}_{3,\text{out}}^\dagger(\omega)
	\end{pmatrix}=\bm{U}\begin{pmatrix}
				\hat{a}_{1,\text{in}}^\dagger(\omega)\\ \hat{a}_{2,\text{in}}^\dagger(\omega) 
				\\ \hat{a}_{3,\text{in}}^\dagger(\omega)
			\end{pmatrix}\ ,
\end{equation}
where $\bm{U}$ must be a $3\times 3$ matrix and is treated here as being frequency independent.
For photon-number-conserving scattering, $\bm{U}$ is now a $3\times 3$ unitary matrix with determinant $\text{e}^{i\xi}$.
Thus, the unitary matrix can be expressed as
\begin{equation}
\bm{U}=R(\Omega)\cdot
		\begin{pmatrix}
		\text{e}^{i\xi/3} &0 &0\\
		0	& \text{e}^{i\xi/3} &0\\
		0	& 0& \text{e}^{i\xi/3}
		\end{pmatrix}
\end{equation}
with $R(\Omega)$ now a special unitary $3\times 3$ matrix, i.e., an SU(3) matrix.
In contrast to Sec.~\ref{sec:from2photons},
$\Omega$ now labels the parameter element of SU(3).  

In fact $\Omega$ is an 8-tuple of angles, as 
the matrix $R(\Omega)$ can be written as the product ~\cite{KdG10}
\begin{align}\label{su3}
	R(\Omega)\equiv& T_{23}(\alpha_1,\beta_1,-\alpha_1)T_{12}(\alpha_{2},\beta_2,-\alpha_2)
						\nonumber\\
		&\times T_{23}(\alpha_3,\beta_3,-\alpha_3) \Phi(\gamma_1,\gamma_2)
\end{align}
with
\begin{equation}
	\Omega=\left(\alpha_1,\beta_1,\alpha_2,\beta_2,\alpha_3,\beta_3,\gamma_1,\gamma_2\right)
\end{equation}
the octuple of SU(3) Euler-like angles.
The set $\{T_{ij}\}$ comprises  SU(2) subgroup matrices
\begin{equation}
\label{eq:r23}
	T_{23}(\alpha,\beta,\gamma)
		=\begin{pmatrix}1&0&0\\0& \text{e}^{-\frac{1}{2}\text{i}(\alpha+\gamma )}\cos\frac{\beta}{2}
			&-\text{e}^{-\frac{1}{2}\text{i}(\alpha -\gamma )}\sin\frac{\beta}{2} \\
			0 & \text{e}^{\frac{1}{2}\text{i}(\alpha -\gamma )} \sin \frac{\beta}{2} & 
			\text{e}^{\frac{1}{2}\text{i}(\alpha +\gamma )} \cos \frac{\beta}{2} 
		\end{pmatrix}
\end{equation}
or
\begin{equation}
	T_{12}(\alpha,\beta,\gamma)
		=\begin{pmatrix}\text{e}^{-\frac{1}{2}\text{i}(\alpha+\gamma )}\cos \frac{\beta}{2}
			&-\text{e}^{-\frac{1}{2}\text{i}(\alpha-\gamma)}\sin\frac{\beta}{2} 
			& 0\\ \text{e}^{\frac{1}{2}\text{i}(\alpha -\gamma )} \sin \frac{\beta}{2}
			&\text{e}^{\frac{1}{2}\text{i}(\alpha +\gamma )}\cos\frac{\beta}{2} &0\\0&0&1
		\end{pmatrix},
\label{eq:r12}
\end{equation}
depending on the values of $(ij)$.
Also 
\begin{equation}
	\Phi(\gamma_1,\gamma_2)
		=\text{diag}(\text{e}^{-2\text{i}\gamma_1},\text{e}^{\text{i}(\gamma_1-\gamma_2/2)},
			\text{e}^{\text{i}(\gamma_1+\gamma_2/2)}).
\label{diagonalh}
\end{equation}
Factorizing Eq.~(\ref{su3}) into SU(2) submatrices corresponds physically to a sequence 
of SU(2) phase-shifter/beamsplitter/phase-shifter transformations on modes (23), (12), and (23), with SU(2) parameters defined by the Euler angles~\cite{YMK86}.

\subsection{Wigner $D$ functions and representations}
\label{subsec:Dreps}

Representations of SU(3) are labeled by two non-negative integers $(p,q)
$~\cite{BB63,Sla81}. This two-integer labeling is a natural extension 
from the SU(2) labeling of representations by a single non-negative 
integer~$2j$, which is the total photon number in the two-mode case and is 
thus analogous to twice the angular momentum. The $(p,q)$ labels are 
defined explicitly in Eq.~(\ref{eq:pq}), but the essence of this 
definition is that, for three photons entering three ports, outputs depend 
on interference between various outcomes (generalizing the idea that the 
Hong-Ou-Mandel dip is due to destructive interference between both photons 
being transmitted and both being reflected as discussed at the end of 
Sec.~\ref{sec:from2photons}). These inferences are accounted for by 
considering how to partition the cases where three photons are divided 
according to a partition $[\lambda_1,\lambda_2,\lambda_3]$ into three 
output ports such that
\begin{equation}
	\lambda_1+\lambda_2+\lambda_3=3.\label{eq:lambdasum}
\end{equation}
Only the difference between total photon numbers in the three partitions 
are needed so just the pair~$(p,q)$ defined in Eq.~(\ref{eq:pq}) is 
needed, not a triple; hence $(p,q)$ serves as a good labeling for states 
corresponding to partitioning photons into channels.

The $3\times 3$ matrices of the form given in Eq.~(\ref{su3}) carry the SU(3) irrep $(1,0)$.
The generalization to SU(3) of Eq.~(\ref{eq:basicsu2matrix}) is thus
\begin{equation}
\label{eq:fundamentalrep}
	R=\begin{pmatrix}
			D^{(1,0)}_{(100),(100)} & D^{(1,0)}_{(100),(010)} & D^{(1,0)}_{(100),(001)} \\
			D^{(1,0)}_{(010),(100)} & D^{(1,0)}_{(010),(010)} & D^{(1,0)}_{(010),(001)}\\
			D^{(1,0)}_{(001),(100)} & D^{(1,0)}_{(001),(010)} & D^{(1,0)}_{(001),(001)}
\end{pmatrix},
\end{equation}
with dependence of $R$ and $D$ on $\Omega$ implicit. 

An expression for the matrix entries in Eq.~(\ref{eq:fundamentalrep})
is easily obtained by explicit multiplication of the matrices of Eqs.~(\ref{eq:r23})-(\ref{diagonalh}) per the sequence of Eq.~(\ref{su3}); e.g.,
\begin{equation}
	D^{(1,0)}_{(010),(100)}
		=\text{e}^{i (\alpha_2-2 \gamma_1)} \cos\frac{\beta_1}{2} \sin\frac{\beta_2}{2}.
\end{equation}
In $D^{(1,0)}_{\bm{\nu},\bm{n}}$, 
the triple
\begin{equation}
\label{eq:outputtriple}
	\bm{\nu}:= (\nu_1\nu_2\nu_3)
\end{equation}
is the occupancy of the output state in channels (1,2,3), and
\begin{equation}
\label{eq:inputtriple}
	\bm{n}:= (n_1n_2n_3)
\end{equation}
is the occupancy of the input channel.

Consequently, for specified $\Omega$,
$D^{(1,0)}_{(010),(100)}$ is the amplitude for scattering with one photon entering 
port~1 and exiting port~2.
The input state
\begin{equation}
	\rket{1(\omega_1)1(\omega_2)1(\omega_3)}^\text{S}
		:=\hat a_{1,\text{in}}^\dagger(\omega_1)\hat a_{2,\text{in}}^\dagger(\omega_2)
		\hat a_{3,\text{in}}^\dagger(\omega_3)\ket{0}
\end{equation}
can scatter to $3^3=27$ possible output states:
\begin{align}
	U\rket{1 (\omega_1)&1(\omega_2)1(\omega_3)}^\text{S}\nonumber\\
		=& [U\hat a_{1,\text{in}}^\dagger(\omega_1)][U\hat a_{2,\text{in}}^\dagger(\omega_2)]
		[U\hat a_{3,\text{in}}^\dagger(\omega_3)]\ket{0}.\nonumber
\end{align}
If the output state is one of the six possible states 
containing photons in distinct ports (here post selected for outputs in ports $i$, $j$, and $k$),
\begin{equation}
	\hat a_{i,\text{out}}^\dagger(\omega_1)\hat a_{j,\text{out}}^\dagger(\omega_2)\hat a_{k,\text{out}}^\dagger(\omega_3)\ket{0},
	i\ne j\ne k\ne i,
\end{equation}
then the amplitude for scattering from the initial to this final state is, up to a constant overall phase  $\text{e}^{i\xi}$,
given by
\begin{equation}
	D^{(1,0)}_{i,(100)}D^{(1,0)}_{j,(010)}D^{(1,0)}_{k,(001)}
\label{eq:su3productfunctions}
\end{equation}
where $\text{e}^{i\xi/3}D^{(1,0)}_{i,(100)}$
denotes the scattering amplitude
\begin{equation}
\label{eq:scatteringamplitude}
	U\hat{a}_{1,\text{in}}^\dagger(\omega_1)\ket{0}\to\hat{a}_{i,\text{out}}^\dagger(\omega_1)\ket{0}.
\end{equation}

To avoid repetitions of products like $$D^{(1,0)}_{i,(100)}D^{(1,0)}_{j,(010)}D^{(1,0)}_{k,(001)},$$
we introduce the shorthand~$R_{ij}$ as the entry $(i,j)$ in the unitary matrix $R$ of Eq.~(\ref{eq:fundamentalrep})
and introduce a shorthand notation:
\begin{equation}
\label{eq:aijk}
	a_{ijk}:= R_{i1}R_{j2}R_{k3}.
\end{equation}
Thus, for instance,
\begin{equation}
	a_{231}
		=R_{21}R_{32}R_{13}
		=D^{(1,0)}_{(010),(100)}D^{(1,0)}_{(001),(010)}D^{(1,0)}_{(100),(001)}.
\end{equation}

Products of the type~(\ref{eq:su3productfunctions})
can be expanded in terms of SU(3) $D^{(p,q)}_{\bm{\nu},\bm{n}}$-functions for higher representations [compare Eq.~(\ref{D1asasum})].
Which values $(p,q)$ to use in the expansion of Eq.~(\ref{eq:su3productfunctions}) can be determined as follows.

Because each monochromatic photon state
$$
\hat a_{i,\text{out}}^\dagger(\omega)\vert 0\rangle
$$
is a basis state for the (three-dimensional) representation $(1,0)$, with the SU(3) scattering matrix  
given in Eq.~(\ref{eq:fundamentalrep}), the product of three photon states is an element in the Hilbert space that carries
the tensor product $(1,0)\otimes(1,0)\otimes (1,0)$ of SU(3).  
This Hilbert space decomposes into the sum of SU(3) irreps given by~\cite{BB63,Spe63,Lic78,ORe82}
\begin{widetext}
\setcounter{MaxMatrixCols}{13}
\begin{equation}
\label{eq:SU3decomposition}
	\begin{matrix}
	(1,0)
		&\otimes	&(1,0)
		& \otimes	&(1,0)
		&\to		&(3,0)
		&\oplus 	&(1,1)
		&\oplus	&(1,1)
		&\oplus	&(0,0) \\
			\Yboxdim{6pt}\yng(1)
		&\otimes	&\Yboxdim{6pt}\yng(1)
		&\otimes	&\Yboxdim{6pt}\yng(1)
		&\to		&\Yboxdim{6pt}\yng(3)
		&\oplus 	&\Yboxdim{6pt}\yng(2,1)
		&\oplus	&\Yboxdim{6pt}\yng(2,1)
		&\oplus	&\Yboxdim{6pt}\yng(1,1,1)
\end{matrix}
\end{equation}
\end{widetext}
where, in addition to the labeling of SU(3) irreps by non-negative integers $(p,q)$, we also provide the labeling and
decomposition in terms of Young diagrams.
The connection between the partition $[\lambda_1,\lambda_2,\lambda_3]$ such that the sum~(\ref{eq:lambdasum}) holds
and
\begin{equation}
	\lambda_1\ge\lambda_2\ge\lambda_3,
\end{equation}
the Young diagram containing $\lambda_i$ boxes on row $i$ and the labels $(p,q)$ and is simple:
\begin{equation}
\label{eq:pq}
	p:=\lambda_1-\lambda_2,\;q{ :}=\lambda_2-\lambda_3.
\end{equation}

From this decomposition we infer that, in general, only functions with $(p,q)=(3,0)$,$(1,1)$ or $(0,0)$ can occur, so that 
\begin{align}
\label{eq:productofSU3Ddecomposed}
	a_{ijk}
		:=&D^{(1,0)}_{i,(100)}D^{(1,0)}_{j,(010)}D^{(1,0)}_{k,(001)}
					\nonumber\\
		=&c^{\Yboxdim{4pt}\yng(3)}_{ijk}D^{\Yboxdim{4pt}\yng(3)}_{(111)1;(111)1}
		+c^{\Yboxdim{4pt}\yng(1,1,1)}_{ijk}D^{\Yboxdim{4pt}\yng(1,1,1)}_{(111)0;(111)0}
					\nonumber\\&
		+c^{\Yboxdim{4pt}\yng(2,1)}_{ijk,(11)}D^{\Yboxdim{4pt}\yng(2,1)}_{(111)1;(111)1}+c^{\Yboxdim{4pt}\yng(2,1)}_{ijk,(00)}D^{\Yboxdim{4pt}\yng(2,1)}_{(111)0;(111)0}
					\nonumber\\&
		+c^{\Yboxdim{4pt}\yng(2,1)}_{ijk,(10)}D^{\Yboxdim{4pt}\yng(2,1)}_{(111)1;(111)0}
					\nonumber\\&
		+c^{\Yboxdim{4pt}\yng(2,1)}_{ijk,(01)}D^{\Yboxdim{4pt}\yng(2,1)}_{(111)0;(111)1},
\end{align}
where, for  later convenience, Young diagrams are used to label all SU(3) irrep except the $(1,0)$ case,
which does not appear in Eq.~(\ref{eq:productofSU3Ddecomposed}) anyway.
We employ standard expressions for the $$D^{(p,q)}_{\bm{\nu}I,\bm{n}J}$$ functions of the irrep $(p,q)$ and notation~\cite{RSdG99}.
The extra indices~$I$ and~$J$, which are not strictly required for labeling states of $(1,0)$,
are used to refer to the transformation properties of the output and input states, respectively,
under the SU(2) subgroup of matrices of type $T_{23}(\alpha,\beta,\gamma)$ given in Eq.~(\ref{eq:r23}).

Table \ref{table:tableofcijk} lists the expansion coefficient of Eq.~(\ref{eq:productofSU3Ddecomposed}) needed to decompose various
relevant triple products of $D^{(1,0)}$ functions. 
The various $c$ coefficients can be obtained by using 
Clebsch-Gordan techniques or by comparing the explicit expressions of the $D$ functions on 
the left-hand side and right-hand side of Eq.~(\ref{eq:productofSU3Ddecomposed}).
\begin{table}[h]
\caption{Coefficients occurring in the expansion of Eq.~(\ref{eq:productofSU3Ddecomposed}).}
\label{table:tableofcijk}
{\renewcommand{\arraystretch}{1.6}
\begin{tabular}{|l|c||c|c|c|c||c|}
\hline
$(ijk)$&$c^{\Yboxdim{4pt}\yng(3)}_{ijk}$
&$c^{\Yboxdim{4pt}\yng(2,1)}_{ijk,(11)}$&$c^{\Yboxdim{4pt}\yng(2,1)}_{ijk,(00)}$
&$c^{\Yboxdim{4pt}\yng(2,1)}_{ijk,(10)}$&$c^{\Yboxdim{4pt}\yng(2,1)}_{ijk,(01)}$&$c^{\Yboxdim{4pt}\yng(1,1,1)}_{ijk}$\\ 
\hline
(123)&$\frac{1}{6}$&$\frac{1}{3}$&$\frac{1}{3}$&0&0&$\frac{1}{6}$\\ \hline
(132)&$\frac{1}{6}$&$\frac{1}{3}$&$-\frac{1}{3}$&0&0&$-\frac{1}{6}$\\ \hline 
(213)&$\frac{1}{6}$&$-\frac{1}{6}$&$\frac{1}{6}$&$\frac{1}{2\sqrt{3}}$&$\frac{1}{2\sqrt{3}}$&$-\frac{1}{6}$\\ \hline 
(231)&$\frac{1}{6}$&$-\frac{1}{6}$&$-\frac{1}{6}$&$-\frac{1}{2\sqrt{3}}$&$\frac{1}{2\sqrt{3}}$&$\frac{1}{6}$\\ \hline 
(312)&$\frac{1}{6}$&$-\frac{1}{6}$&$-\frac{1}{6}$&$\frac{1}{2\sqrt{3}}$&$-\frac{1}{2\sqrt{3}}$&$\frac{1}{6}$\\ \hline
(321)&$\frac{1}{6}$&$-\frac{1}{6}$&$\frac{1}{6}$&$-\frac{1}{2\sqrt{3}}$&$-\frac{1}{2\sqrt{3}}$&$-\frac{1}{6}$ \\ \hline
\end{tabular}}
\end{table}

Thus, for $(ijk)=(132)$, we have
\begin{align}
	a_{132}
		=& D^{(1,0)}_{(100),(100)}D^{(1,0)}_{(001),(010)}D^{(1,0)}_{(010),(001)} \nonumber \\
		=&\frac{1}{6}D^{\Yboxdim{4pt}\yng(3)}_{(111)1;(111);1} - \frac{1}{6}D^{\Yboxdim{4pt}\yng(1,1,1)}_{(111)0;(111);0}\nonumber \\
		&+\frac{1}{3}D^{\Yboxdim{4pt}\yng(2,1)}_{(111)1;(111);1}-\frac{1}{3}D^{\Yboxdim{4pt}\yng(2,1)}_{(111)0;(111);0}.
\end{align}

The SU$(3)$ irrep {\tiny\Yvcentermath1$\yng(2,1)$} occurs twice in Eq.~(\ref{eq:SU3decomposition}).
The two copies of the {\tiny\Yvcentermath1$\yng(2,1)$} representation are mathematically indistinguishable, although the
states in each representation are distinct.
Note that even if the states are in different copies of {\tiny\Yvcentermath1$\yng(2,1)$}, 
the $D^{\Yboxdim{4pt}\yng(2,1)}$ functions are identical.  Further discussion and examples are given in Appendix~\ref{appendixsu3}.
A similar situation occurs in treating a system comprising three spin-1/2 particles:
the final set of states contains two distinct sets of $s=1/2$; although the states in the sets are distinct, both sets transform as 
$s=1/2$ objects.

\subsection{$S_3$, partitions and immanants}
\label{subsec:S3partitionsimmanants}

In addition to labelling SU(3) irreps, the Young diagrams of Eq.~(\ref{eq:SU3decomposition}), namely
\begin{equation}
\label{eq:S3}
	\Yboxdim{6pt}\yng(3),\;\yng(2,1),\;\yng(1,1,1),
\end{equation}
also label the representations of $S_3$, 
which is the six-element permutation group of three objects.  
The permutation group $S_3$ has three irreducible representations: two are of dimension 1 and one is of dimension 2.
Certain matrix functions called immanants are constructed from the entries of a $3\times 3$ matrix using elements in $S_3$
and their irrep characters.
(The characters of a representation
are the traces of the matrix representing elements in the group.
Characters are fundamental to representation theory~\cite{Lit50,DDJ07}.)

For $S_3$ there are three immanants: the permanent, the determinant, 
and another immanant (the permanent and the determinant are special cases of immanants). 
Because specific immanants
are constructed using characters of a specific irrep of $S_3$ denoted by a Young diagram, this Young diagram can
also represent the corresponding immanant.
Table~\ref{table:char} is the character table of $S_3$. The values in this table are required to construct the permanent, immanant, and determinant of a $3\times 3$ matrix~\cite{Lit50}, respectively.

\begin{table}[h!]
\centering
\begin{tabular}{|c|c|c|c|c|}
\hline
Elements &$\mathds{1}$	& $\sigma_{ab}$&$\sigma_{abc}$&\\ 
		&			&	$=\{P_{12},P_{13},P_{23}\}$ & $=\{P_{123},P_{132}\}$&	 \\
\hline
irrep $\lambda$ & $\chi^\lambda(\mathds{1})$ & $\chi^\lambda(\sigma_{ab})$\phantom {\Yvcentermath1\Yboxdim{8pt}$\yng(1,1)$} 
&$\chi^\lambda(\sigma_{abc})$ 
& dim. \\
\hline
\Yboxdim{6pt}\yng(3)      & 1 & 1 &1 & 1\\
\Yboxdim{6pt}\yng(2,1)   & 2 & 0 & -1 & 2 \\
\Yboxdim{6pt}\yng(1,1,1)& 1 &-1 &1 &1 \\
\hline
\end{tabular}\ 
\caption{Character table for $S_3$~\cite{Lit50}.} 
\label{table:char}
\end{table}

One immanant exists for each irrep of $S_3$.  
An immanant of a $3\times 3$ matrix $X:=\left(x_{ij}\right)$,
with~$x_{ij}$ the entry in the $i^\text{th}$ row and $j^\text{th}$ column of~$X$,
is~\cite{Wyb70}
\begin{equation}
\label{eq:Imm}
	\text{imm}^\lambda X:
		=\sum_{\sigma}\chi^\lambda(\sigma)P_\sigma (x_{11}x_{22}x_{33}).
\end{equation}
Here~$\chi^\lambda(\sigma)$ denotes the character of the element $\sigma\in S_n$
for irrep~$\lambda$,
and
\begin{equation}
\label{eq:Psigma}
	P_\sigma (x_{1j}x_{2k}x_{3\ell})=x_{1,\sigma(j)}x_{2,\sigma(k)}x_{3,\sigma(\ell)}
\end{equation}
exchanges entry $x_{aj}$ with entry $x_{a,\sigma(j)}$ where $\sigma(j)$ is the image of $j$
under the element $P_{\sigma}$ of $S_3$.

As
\begin{equation}
	\chi^{\Yboxdim{4pt}\yng(3)}(P_{\sigma})\equiv 1\;\forall\sigma\in S_3,
\end{equation}
the permanent, which corresponds to the Young diagram $\Yboxdim{6pt}\yng(3)$,
is obtained from Eq.~(\ref{eq:Imm}) and yields
\begin{align}
	\text{per}X
		:= &\Yboxdim{6pt}\yng(3)(X)
				\nonumber\\
		=& x_{11} x_{22} x_{33}+x_{11} x_{23} x_{32}
				\nonumber\\ &
			+x_{12} x_{21} x_{33}+ x_{12} x_{23} x_{31}
				\nonumber\\ &
			+x_{13}x_{21} x_{32}+x_{13} x_{22} x_{31}.
\end{align}
The determinant corresponds to the Young diagram {\Yvcentermath1\Yboxdim{4pt}$\yng(1,1,1)$} and is simply
\begin{align}
	\text{det}X
		=&\Yboxdim{4pt}\yng(1,1,1)(X)
				\nonumber\\
		=&x_{11} x_{22} x_{33}-x_{11} x_{23} x_{32}
				\nonumber\\ &
			-x_{12} x_{21} x_{33}+x_{12} x_{23} x_{31}
				\nonumber\\ &
			+x_{13}x_{21} x_{32}-x_{13} x_{22} x_{31}.
\end{align}
Finally, the mixed-symmetry immanant, associated with the Young diagram {\Yvcentermath1\Yboxdim{4pt}$\yng(2,1)$}, is given by
\begin{align}
	\Yboxdim{4pt}\yng(2,1)(X)
		=&2\times \mathds{1} (x_{11}x_{22}x_{33})
				\nonumber\\ &
		+0\times \left(P_{12} +P_{13}+P_{23}\right)
		\left(x_{11}x_{22}x_{33}\right)
				\nonumber\\ &
		-1\times \left(P_{123}+P_{132}\right)(x_{11}x_{22}x_{33})
			\nonumber\\
		=&2x_{11} x_{22} x_{33}-x_{12}x_{23}x_{31}-x_{13}x_{21}x_{32}.
\end{align}
As this is the only immanant for SU(3) that is neither a permanent nor a determinant,
we refer to this intermediate immanant as ``the immanant" and denote this immanant of $X$ by $\text{imm}X$.

\subsection{ $D$ functions and immanants}
\label{subsec:Dfunctionsimmanants}

In this subsection we reprise our earlier observations that link immanants of matrices to $D$-functions for SU(3)~\cite{TGdGS13}.
Then we extend this work with Eqs.~(\ref{eq:immD}) and~(\ref{eq:detT}) being new results.

Generalizing Eq.~(\ref{eq:R21matrix}), we denote by $R^{ijk}$ the matrix obtained from $R$ in 
Eq.~(\ref{eq:fundamentalrep}) by permuting rows
of $R$.
The permutation is done such that the first row of~$R$ becomes row~$i$ of $R^{ijk}$,
the second row of $R$ becomes row~$j$ of $R^{ijk}$,
and the third row of~$R$ becomes row~$k$ of $R^{ijk}$.
The rows of~$R$ and of~$R^{ijk}$ are thus related by the permutation
\begin{equation}
	P^{ijk}(123): (123)\to (ijk).
\end{equation}  
Then, with reference to the coefficients of table \ref{table:tableofcijk},
the following key results for the permanent, the immanant and the determinant
can be verified from the explicit expressions of the SU(3) $D$ functions supplied in the appendices.
 
The permanent of $R^{ijk}$, which we denote by
${\Yboxdim{6pt}\Yvcentermath1\yng(3)}_{\,ijk}$,  is
\begin{equation}
\label{eq:perR}
	\text{per}R^{ijk}={\Yboxdim{6pt}\Yvcentermath1\yng(3)}_{\,ijk}= 
	6 c^{\Yboxdim{4pt}\yng(3)}_{ijk}D^{\Yboxdim{4pt}\yng(3)}_{(111)1;(111)1}.
\end{equation}
The immanant of $R^{ijk}$, 
which we denote by
${\Yboxdim{6pt}\Yvcentermath1\yng(2,1)}_{\,ijk}$,
 is
\begin{align}
	\text{imm}R^{ijk}
		=&{\Yboxdim{6pt}\Yvcentermath1\yng(2,1)}_{\,ijk}\nonumber\\
		=&3\Big(c^{\Yboxdim{4pt}\yng(2,1)}_{ijk,(11)}D^{\Yboxdim{4pt}\yng(2,1)}_{(111)1;(111)1}
				\nonumber\\&
			+c^{\Yboxdim{4pt}\yng(2,1)}_{ijk,(00)}D^{\Yboxdim{4pt}\yng(2,1)}_{(111)0;(111)0}
				\nonumber\\&
			+c^{\Yboxdim{4pt}\yng(2,1)}_{ijk,(10)}D^{\Yboxdim{4pt}\yng(2,1)}_{(111)1;(111)0}
				\nonumber\\&
			+c^{\Yboxdim{4pt}\yng(2,1)}_{ijk,(01)}D^{\Yboxdim{4pt}\yng(2,1)}_{(111)0;(111)1}\Big). 
\end{align}
In particular, using the expression for $c^{\Yboxdim{4pt}\yng(2,1)}_{ijk}$,
we obtain 
\begin{align}
	\text{imm}R^{231}
		=&-\text{imm}R^{123}-\text{imm}R^{312},\nonumber\\
	\text{imm}R^{321}
		=&-\text{imm}R^{213}-\text{imm}R^{132},
\label{immdependent}
\end{align}
thereby showing that there are only four linearly independent immanants.
Conversely, it is possible to express the various $D^{\Yboxdim{4pt}\yng(2,1)}_{(111)I;(111)J}$ in terms of the immanants:
\begin{align}
	&\begin{pmatrix}
		D^{\Yboxdim{4pt}\yng(2,1)}_{(111)1;(111)1}\\
		D^{\Yboxdim{4pt}\yng(2,1)}_{(111)0;(111)0}\\
		D^{\Yboxdim{4pt}\yng(2,1)}_{(111)1;(111)0}\\
		D^{\Yboxdim{4pt}\yng(2,1)}_{(111)0;(111)1}
	\end{pmatrix}\nonumber \\ 
	&=\frac{1}{2}
	\begin{pmatrix}
		1 & 0 & 1 & 0 \\
		1 & 0 & -1 & 0 \\
		\frac{1}{\sqrt{3}} & \frac{2}{\sqrt{3}} & \frac{1}{\sqrt{3}} & \frac{2}{\sqrt{3}} \\
		-\frac{1}{\sqrt{3}} & \frac{2}{\sqrt{3}} & \frac{1}{\sqrt{3}} & -\frac{2}{\sqrt{3}}
	\end{pmatrix}
	\begin{pmatrix}
		{\Yboxdim{6pt}\yng(2,1)}_{123}\\
		{\Yboxdim{6pt}\yng(2,1)}_{213}\\
		{\Yboxdim{6pt}\yng(2,1)}_{132}\\
		{\Yboxdim{6pt}\yng(2,1)}_{312}
	\end{pmatrix}.
\label{eq:immD}
\end{align}
The determinant of $R^{ijk}$, which we denote by
${\Yboxdim{6pt}\Yvcentermath1\yng(1,1,1)}_{\,ijk}$, is
\begin{equation}
\label{eq:detT}
	\text{det}R^{ijk}={\Yboxdim{6pt}\Yvcentermath1\yng(1,1,1)}_{\,ijk} =
	6c^{\Yboxdim{4pt}\yng(1,1,1)}_{ijk}D^{\Yboxdim{4pt}\yng(1,1,1)}_{(111)0;(111)0}.
\end{equation}

\subsection{ Amplitudes and immanants}
\label{subsec:amplitudesimmanants}

Using the relations between immanants and $D$ functions in the previous subsection, we
see that the amplitude in Eq.~(\ref{eq:productofSU3Ddecomposed}) can also be written as a linear combination of the immanants, the permanent, and the determinant.
Using the shorthand notation,~(\ref{eq:aijk}), we obtain
\begin{equation}
\label{eq:aM}
	\begin{pmatrix}
		a_{123} \\ 
		a_{132} \\
		a_{213} \\
		a_{231} \\ 
		a_{312} \\ 
		a_{321} \\ 
	\end{pmatrix}
	=M
	\begin{pmatrix}
		{\Yboxdim{6pt}\yng(3)}_{123}\\ 
		{\Yboxdim{6pt}\yng(2,1)}_{123}\\
		{\Yboxdim{6pt}\yng(2,1)}_{132}\\
		{\Yboxdim{6pt}\yng(2,1)}_{213}\\
		{\Yboxdim{6pt}\yng(2,1)}_{312}\\ 
		{\Yboxdim{6pt}\yng(1,1,1)}_{123}
	\end{pmatrix}
\end{equation}
for
\begin{equation}
\label{eq:matrixM}
	M=
	\begin{pmatrix}
		\frac{1}{6} & \frac{1}{3} & 0 & 0 & 0 & \frac{1}{6} \\
		\frac{1}{6} & 0 & \frac{1}{3} & 0 & 0 & -\frac{1}{6} \\
		\frac{1}{6} & 0 & 0 & \frac{1}{3} & 0 & -\frac{1}{6} \\
		\frac{1}{6} & -\frac{1}{3} & 0 & 0 & -\frac{1}{3} & \frac{1}{6} \\
		\frac{1}{6} & 0 & 0 & 0 & \frac{1}{3} & \frac{1}{6} \\
		\frac{1}{6} & 0 & -\frac{1}{3} & -\frac{1}{3} & 0 & -\frac{1}{6}
	\end{pmatrix}
\end{equation}

Finally, in view of Eqs.~(\ref{immdependent}) and (\ref{eq:aM}),
the connection with amplitudes for monochromatic input states is neatly summarized 
by
\begin{align}
^S\big (1(\omega_i)1(\omega_j)1(\omega_k)\big|R
		\rket{1(\omega_1)1(\omega_2)1(\omega_3)}^S \nonumber \\
	=\frac{1}{6}{\Yboxdim{6pt}\Yvcentermath1\yng(3)}_{ijk}
+\frac{1}{3}{\Yboxdim{6pt}\Yvcentermath1 \yng(2,1)}_{ijk}
+\frac{1}{6}{\Yboxdim{6pt}\Yvcentermath1 \yng(1,1,1)}_{ijk},
\label{eq:weightedsum}
\end{align}
where ${\Yboxdim{6pt}\Yvcentermath1\yng(3)}_{ijk}$ is the permanent of the matrix $R^{ijk}$,
${\Yboxdim{6pt}\Yvcentermath1 \yng(1,1,1)}_{ijk}$ is the determinant of the matrix $R^{ijk}$, and
${\Yboxdim{6pt}\Yvcentermath1 \yng(2,1)}_{ijk}$ is the immanant of the matrix $R^{ijk}$.

Equation~(\ref{eq:weightedsum}) is an elegant connection between amplitudes and immanants 
for the special case of monochromatic photon inputs. It generalizes
the analogous result of Eq.~(\ref{su2covariantrate}) in the two-photon case.
These relations can be verified from the explicit expressions of the SU(3) $D$ functions supplied in the Appendices.

We observe that Eq.~(\ref{eq:weightedsum}) is surprisingly simple.
The amplitude is a product of $D^{(1,0)}$ functions, 
and this product
decomposes into a nontrivial sum of $D^{(p,q)}$ functions, which themselves are nontrivial linear combinations 
of immanants.
In particular, it is surprising that a single ${\Yboxdim{6pt}\Yvcentermath1 \yng(2,1)}_{ijk}$ immanant should appear.  
We note that
the coefficients of ${\Yboxdim{6pt}\Yvcentermath1\yng(3)}_{ijk}$ and 
${\Yboxdim{6pt}\Yvcentermath1\yng(1,1,1)}_{ijk}$ are identical,
and the coefficient of ${\Yboxdim{6pt}\Yvcentermath1\yng(2,1)}_{ijk}$ is twice that of 
${\Yboxdim{6pt}\Yvcentermath1\yng(3)}_{ijk}$.
The proportions $1:1:2$ are also the proportions of the dimension of the respective irreps of $S_3$.

\section{Three-photon coincidence and immanants}\label{obs}

In this section we develop the general formula for three-photon coincidence rate~$\wp$
given one photon entering each input port of a passive three-channel optical interferometer at arbitrary times~$\bm{\tau}$~(\ref{eq:tau}).
In Sec.~\ref{subsec:general} we introduce the formalism for the general input and resultant output state
and the consequent formula for the coincidence rate.
Then Sec.~\ref{subsec:simultaneity} focuses on the special case where all photons are simultaneous,
i.e., where $\bm{\tau}\equiv(\tau,\tau,\tau)$.
This case of no delays is the case normally assumed in the literature on BosonSampling.

The case where two photons arrive simultaneously and one either precedes or follows those two by a significant time duration
is the topic of Sec.~\ref{subsec:twoandone}.
This subsection probes the Hong-Ou-Mandel dip limit where two photons can exhibit a dip 
given the right choice of~$\Omega$ and a third photon is independent.
Finally, in Sec.~\ref{subsec:distinguishable}
we deal with the case where the photon arrival times are far apart but yield coincidences because the detector integration time is 
of course sufficiently long.

\subsection{General case}
\label{subsec:general}

A three-photon input state with general spectral profile, 
but identical for each of the three incoming modes,
is written as
\begin{align}
\label{eq:syminput}
	\ket{\psi}_\text{in}
		=&\int\text{d}^3\bm{\omega}
			\text{e}^{-\text{i}\bm{\omega}\cdot\bm{\tau}}\tilde{\phi}(\omega_1)\tilde{\phi}(\omega_2)\tilde{\phi}(\omega_3)
				\nonumber\\&\times
			\hat{a}_1^\dag(\omega_1)\hat{a}_2^\dag(\omega_2)\hat{a}_3^\dag(\omega_3) \ket{0}
\end{align}
for $\bm{\omega}:=(\omega_1,\omega_2,\omega_3)$
the three-dimensional frequency and d$^3\bm{\omega}$ the three-dimensional measure over this domain.
The exponential involves the dot product between $\bm\omega$,
and the three-vector time-of-entry vector~$\bm\tau$ for the photons,~(\ref{eq:tau}).

Passage through the interferometer produces
\begin{align}
	\ket{\psi}_{\rm out}
		=&\int\text{d}^3\bm{\omega}\text{e}^{-\text{i}\bm{\omega}\cdot\bm{\tau}}
		\tilde{\phi}(\omega_1)\tilde{\phi}(\omega_2)\tilde{\phi}(\omega_3)
			\nonumber\\
	&\times\left(U\hat{a}_1^\dag(\omega_1)\right)
	\left(U\hat{a}_2^\dag(\omega_2)\right)\left(U\hat{a}_3^\dag(\omega_3)\right) \ket{0}.
\end{align}
The coincidence rate depends only on pairwise time delays given by the two-component vector,~$\bm\Delta$ (\ref{eq:Delta}),
but the expressions for the coincidence rate~$\wp$ are easier to understand in terms of the three-component vector~$\bm\tau$ (\ref{eq:tau}).
Therefore, we express the three-photon coincidence rate in the form
\begin{widetext}
\begin{align}
	\wp(\bm{\Delta};\Omega)
		=&\int\text{d}^3\tilde{\bm{\omega}}
		\left|\phi(\tilde\omega_1)\right|^2
		\left|\phi(\tilde\omega_2)\right|^2
		\left|\phi(\tilde\omega_3)\right|^2
		\Big|\,a_{123}\text{e}^{i\bm{\tilde{\omega}}\cdot\bm{\tau}}
		+a_{132}\text{e}^{\text{i}(\tilde\omega_1\tau_1+\tilde\omega_3\tau_2+\tilde\omega_2\tau_3)}
		+a_{213}\text{e}^{\text{i}(\tilde\omega_2\tau_1+\tilde\omega_1\tau_2+\tilde\omega_3\tau_3)}\nonumber\\
		&+a_{231}\text{e}^{\text{i}(\tilde\omega_2\tau_1+\tilde\omega_3\tau_2+\tilde\omega_1\tau_3)}
		+a_{312}\text{e}^{\text{i}(\tilde\omega_3\tau_1+\tilde\omega_1\tau_2+\tilde\omega_2\tau_3)}
		+a_{321}\text{e}^{\text{i}(\tilde\omega_3\tau_1+\tilde\omega_2\tau_2+\tilde\omega_1\tau_3)}\Big|^2
\label{fullrate3photoncase}
\end{align}
\end{widetext}
where $\tau_1$ can be set to 0 but is kept arbitrary in the explicit expression, and~$\bm\tau$ and~$\bm\Delta$ are related by Eq.~(\ref{eq:tauDelta}).
Each~$a_{ijk}$ can be written in terms of $D^{(p,q)}$ functions per Eq.~(\ref{eq:productofSU3Ddecomposed})
or in terms of immanants per Eq.~(\ref{eq:aM}).
For the explicit dependence of the three-photon coincidence rate in Eq.~(\ref{fullrate3photoncase}) in terms of $D^{(p,q)}$ functions upon integration over the frequencies, please refer to Appendix \ref{rateD}.

\subsection{Simultaneity}
\label{subsec:simultaneity}

We first consider $\wp(\bm{\Delta}=\bm{0};\Omega)$ corresponding to all photons arriving simultaneously, in which case the phases in Eq.~(\ref{fullrate3photoncase}) effectively disappear upon taking the squared modulus.
The sum of  $a_{ijk}$ coefficients is easily evaluated using Eq.~(\ref{eq:aM}) to be the permanent of the matrix.
Therefore, the three-photon coincidence rate
\begin{equation}
\label{eq:per111}
	\wp(\bm{0};\Omega)\propto \vert {\rm per}(\Omega)\vert^2
\end{equation}
adopts a simple form with respect to the octuple~$\Omega$.

The proportionality of the coincidence rate to the squared modulus of the permanent,~(\ref{eq:per111}), is the heart of the BosonSampling Problem
and its interferometrically friendly test~\cite{AA11}.
This case of simultaneity is also the focus of research into the Hong-Ou-Mandel dip extension to three-channel passive optical interferometry~\cite{Cam00}.

\subsection{Two simultaneous photons and one delayed}
\label{subsec:twoandone}

Suppose now that two of the delays are the same, but a third is different in the sense that its arrival time is significantly earlier or
later than when the other two arrive.
This significant delay~$\tau$ corresponds to a duration longer than the photon pulse duration.
In this case we write $\bm{\Delta}=(\tau,0)$.

Photons~$2$ and~$3$ are then simultaneous,
and the input state can be written in the reduced form
\begin{align}
	\ket{111}_{\rm sym}
		=&\frac{1}{2}\int\text{d}^3\bm{\omega}
		\phi(\omega_1)\phi(\omega_2)\phi(\omega_3)
			\nonumber\\ &
		\times\text{e}^{-\text{i}\omega_1\tau_1} \text{e}^{-i (\omega_2+\omega_3)\tau}\hat{a}_1^\dag(\omega_1)
			\nonumber\\ &
		\times\left(\hat{a}_2^\dag(\omega_2)\hat{a}_3^\dag(\omega_3)
		+\hat{a}_2^\dag(\omega_3)\hat{a}_3^\dag(\omega_2)\right)\ket{0},
\end{align}
which is symmetric under exchange of the $2$ and $3$ labels.

The coincidence rate is then given by the expression
\begin{align}
\label{eq:2indenticalphotons}
	\wp(\bm{\Delta};\Omega)
		=&\vert A\vert^2 + \vert B\vert^2+ \vert C\vert^2+\text{e}^{-\sigma^2\tau^2}\big[(A^*+B^*)C
				\nonumber \\ &
			+(A^*+C^*)B+(B^*+C^*)A \big]
\end{align}
where 
the functions $A$, $B$ and $C$  are related to immanants by
\begin{align}
\label{eq:ABC}
A&=a_{123}+a_{132}=\frac{1}{3}\left({\Yboxdim{6pt}\Yvcentermath1\yng(3)}
	+{\Yboxdim{6pt}\Yvcentermath1 \yng(2,1)}_{123}
	+{\Yboxdim{6pt}\Yvcentermath1 \yng(2,1)}_{132} \right),\nonumber\\
B&=a_{213}+a_{231}=\frac{1}{3}\left({\Yboxdim{6pt}\Yvcentermath1\yng(3)}
	+{\Yboxdim{6pt}\Yvcentermath1 \yng(2,1)}_{213}
	+{\Yboxdim{6pt}\Yvcentermath1 \yng(2,1)}_{231} \right),\\
C&=a_{321}+a_{312}=\frac{1}{3}\left({\Yboxdim{6pt}\Yvcentermath1\yng(3)}
	+{\Yboxdim{6pt}\Yvcentermath1 \yng(2,1)}_{312}
	+{\Yboxdim{6pt}\Yvcentermath1 \yng(2,1)}_{321}\right).\nonumber
\end{align}
Alternatively,
$A$, $B$ and $C$  are given in terms of $D^{(p,q)}$ functions by
\begin{align}
\label{eq:23symmetry}
	A=&\frac{1}{3}\left(D^{\Yboxdim{4pt}\Yvcentermath1\yng(3)}_{(111)1;(111)1}
	+2 D^{\Yboxdim{4pt}\Yvcentermath1\yng(2,1)}_{(111)1;(111)1}\right) \nonumber\\
	B=&\frac{1}{3}\left(D^{\Yboxdim{4pt}\Yvcentermath1\yng(3)}_{(111)1;(111)1}
	- D^{\Yboxdim{4pt}\Yvcentermath1\yng(2,1)}_{(111)1;(111)1}\right.\nonumber \\
		&\left.+\sqrt{3}D^{(1,1)}_{(111)0;(111)1}\right)\\
	C=&\frac{1}{3}\left(D^{\Yboxdim{4pt}\Yvcentermath1\yng(3)}_{(111)1;(111)1}
	- D^{\Yboxdim{4pt}\Yvcentermath1\yng(2,1)}_{(111)1;(111)1}\right.\nonumber \\
		&\left. -\sqrt{3}D^{\Yboxdim{4pt}\Yvcentermath1\yng(2,1)}_{(111)0;(111)1}\right).\nonumber
\end{align}

For $\tau\to\infty$, the rate collapses to 
\begin{equation}
	\lim_{\Delta\to\infty}\wp((\Delta,0);\Omega)\to\vert A \vert ^2 + \vert B \vert ^2 + \vert C \vert ^2.
\label{asymptoticP111}
\end{equation}
Further insight into the connection between immanants and $D$-functions is gained by noting that
insertion of expressions of Eqs.~(\ref{eq:ABC}) into Eq.~(\ref{eq:2indenticalphotons}) yields
\begin{align}
	\vert A\vert^2&+ \vert B\vert^2+ \vert C\vert^2
		\nonumber\\
	=&\frac{2}{3}\left[\left\vert D^{\Yboxdim{4pt}\Yvcentermath1\yng(2,1)}_{(111)0;(111)1}\right\vert^2
	+\left\vert D^{\Yboxdim{4pt}\Yvcentermath1\yng(2,1)}_{(111)1;(111)1}\right\vert^2\right]
		\nonumber\\&
	+\frac{1}{3}\left|D^{\Yboxdim{4pt}\Yvcentermath1\yng(3)}_{(111)1;(111)1}\right|^2
\label{sumofsquares}
\end{align}
whereas
\begin{align}
	(A^*+B^*)&C+(A^*+C^*)B+(B^*+C^*)A
				\nonumber\\
		=&-\frac{2}{3}\left[\vert D^{\Yboxdim{4pt}\Yvcentermath1\yng(2,1)}_{(111)0;(111)1}\vert^2
		+\vert D^{\Yboxdim{4pt}\Yvcentermath1\yng(2,1)}_{(111)1;(111)1}\vert^2\right]
				\nonumber\\&
		+\frac{2}{3}\left|D^{\Yboxdim{4pt}\Yvcentermath1\yng(3)}_{(111)1;(111)1}\right|^2.
\label{crossterms}
\end{align}

From Eqs.~(\ref{sumofsquares}) and (\ref{crossterms}),
we see that coincidence measurements with $\Delta_2=0$ only 
yields information only about the sum 
\begin{equation}
	\left\vert D^{\Yboxdim{4pt}\Yvcentermath1\yng(2,1)}_{(111)0;(111)1}\right\vert^2
	+\left\vert D^{\Yboxdim{4pt}\Yvcentermath1\yng(2,1)}_{(111)1;(111)1}\right\vert^2
\end{equation}
and not about $$D^{\Yboxdim{4pt}\Yvcentermath1\yng(2,1)}_{(111)0;(111)1} \text{ or }
D^{\Yboxdim{4pt}\Yvcentermath1\yng(2,1)}_{(111)1;(111)1}$$ separately.

The reason that these specific $D$-functions occur can be understood by observing that the state
\begin{equation}
\label{eq:23triplet}
	\left[\hat{a}_2^\dag(\omega_2)\hat{a}_3^\dag(\omega_3) + 
		\hat{a}_2^\dag(\omega_3)\hat{a}_3^\dag(\omega_2)\right]\ket{0}
\end{equation} 
is obviously symmetric under permutation of the frequencies.
Consequently, the state~(\ref{eq:23triplet}) is also a state
of angular momentum $I_{23}=1$ with the angular momentum label $I_{23}$ referring 
to the subgroup SU(2)$_{23}$ of matrices mixing modes $2$ and $3$, 
as discussed in Appendix \ref{appendixsu3}.
Permutation symmetry explains why only $D$ functions of the type
$$D^{(p,q)}_{(111)I_{23},(111)1}$$ can enter into the rate when $\Delta_2=0$.

Furthermore the state of Eq.~(\ref{eq:23triplet}) belongs to the $(2,0)$ irrep of SU(3).  
The resultant three-photon Hilbert space
is thus the subspace of the full Hilbert space decomposed in Eq.~(\ref{eq:SU3decomposition})
and is now spanned by states in the SU(3) irreps
\begin{equation}\label{10x20}
	\begin{matrix}
	\Yboxdim{6pt}\yng(1)	&\otimes &\Yboxdim{6pt}\yng(2) &\to & \Yboxdim{6pt}\yng(3)&\oplus  &\Yboxdim{6pt}\yng(2,1)\\
	(1,0)                          		&\otimes & (2,0) &\to & (3,0)  & \oplus   & (1,1).
	\end{matrix}
\end{equation}
As a consequence, only functions in the $(3,0)$ and $(1,1)$ irreps can appear in the final rate.  

This symmetry property under exchange of modes
$2$ and $3$ can be made explicit in terms of immanants.
First, note that the right-hand side of
\begin{equation}
	D^{\Yboxdim{4pt}\Yvcentermath1 \yng(2,1)}_{(111)1;(111)1}
		=\frac{1}{2}\left(
		{\Yboxdim{6pt}\Yvcentermath1 \yng(2,1)}_{123}
		+{\Yboxdim{6pt}\Yvcentermath1 \yng(2,1)}_{132}\right)
\end{equation}
is evidently symmetric under exchange of $2$ and $3$.  
The symmetry of 
$$D^{(1,1)}_{(111)0;(111)1}$$ is slightly more delicate.
We start by observing that this function can be written in two ways, namely,
\begin{align}
\label{eq:d01first}
	D^{\Yboxdim{4pt}\Yvcentermath1 \yng(2,1)}_{(111)0;(111)1}
		=&\frac{1}{2\sqrt{3}}\left(
			{\Yboxdim{6pt}\Yvcentermath1 \yng(2,1)}_{123}
				+{\Yboxdim{6pt}\Yvcentermath1 \yng(2,1)}_{132}\right) 
						\nonumber \\ &
		+\frac{1}{\sqrt{3}}
			\left(
				{\Yboxdim{6pt}\Yvcentermath1 \yng(2,1)}_{213}
					+{\Yboxdim{6pt}\Yvcentermath1 \yng(2,1)}_{231}
			\right)
\end{align}
or, alternatively, as
\begin{align}
\label{d01second}
	D^{\Yboxdim{4pt}\Yvcentermath1 \yng(2,1)}_{(111)0;(111)1}
		=&-\frac{1}{2\sqrt{3}}\left(
			{\Yboxdim{6pt}\Yvcentermath1 \yng(2,1)}_{123}
				+{\Yboxdim{6pt}\Yvcentermath1 \yng(2,1)}_{132}\right)
						\nonumber \\ &
		-\frac{1}{\sqrt{3}}
			\left(
				{\Yboxdim{6pt}\Yvcentermath1 \yng(2,1)}_{312}
					+{\Yboxdim{6pt}\Yvcentermath1 \yng(2,1)}_{321}
			\right).
\end{align}
Exchanging the labels $2$ and $3$ in  Eq.~(\ref{eq:d01first}) transforms this expression
into the negative of Eq.~(\ref{d01second}).
In other words, the function $$D^{\Yboxdim{4pt}\Yvcentermath1 \yng(2,1)}_{(111)0;(111)1}$$
is antisymmetric under exchange of output photons $2$ and $3$, as expected
from the $I_{23}=0$ (singlet) nature of the output state.
However, the rate is expressed in terms of the modulus square of the function
so the rate is actually symmetric under exchange of output photons $2$ and $3$.

Now suppose instead that $\bm{\Delta}=(0,\tau)$.
Photons $1$ and $2$ are now identical and
the input state can be written as
\begin{align}
	\ket{111}_{\rm sym}
		=&\frac{1}{2}\int\text{d}^3\bm{\omega}
		\phi(\omega_1)\phi(\omega_2)\phi(\omega_3)
		\hat{a}_3^\dag(\omega_3)
			\nonumber\\&\times
		\left(\hat{a}_1^\dag(\omega_1)\hat{a}_2^\dag(\omega_2)
		+\hat{a}_1^\dag(\omega_2)\hat{a}_2^\dag(\omega_1)\right)\ket{0},
\end{align}
which is symmetric now under exchange of the $1$ and $2$ labels.
The coincidence rate now takes the form of Eq.~(\ref{eq:2indenticalphotons}), but with 
$A,B$ and $C$ now given by
\begin{align}
	A&=a_{123}+a_{213}\nonumber \\
	&=\frac{1}{3}\left({\Yboxdim{6pt}\Yvcentermath1\yng(3)}
	+{\Yboxdim{6pt}\Yvcentermath1 \yng(2,1)}_{123}
	+{\Yboxdim{6pt}\Yvcentermath1 \yng(2,1)}_{213}\right)\ , \label{A12}
 \\
B&=a_{132}+a_{312}\nonumber \\
&=\frac{1}{3}\left( {\Yboxdim{6pt}\Yvcentermath1\yng(3)}
+{\Yboxdim{6pt}\Yvcentermath1 \yng(2,1)}_{132}
+{\Yboxdim{6pt}\Yvcentermath1 \yng(2,1)}_{312} \right)\ ,\label{B12} \\
C&=a_{231}+a_{321}\nonumber \\
&=\frac{1}{3}\left({\Yboxdim{6pt}\Yvcentermath1\yng(3)}
+{\Yboxdim{6pt}\Yvcentermath1 \yng(2,1)}_{231}
+{\Yboxdim{6pt}\Yvcentermath1 \yng(2,1)}_{321}\right).  \label{C12}
\end{align}
Note that $A$, $B$, and $C$ 
are now symmetric under interchange of the first two indices of each term.

The state
\begin{equation}
\left(\hat{a}_1^\dag(\omega_1)\hat{a}_2^\dag(\omega_2)
		+\hat{a}_1^\dag(\omega_2)\hat{a}_2^\dag(\omega_1)\right)\ket{0},
\end{equation}
now has a definite angular momentum $I_{12}=1$,
where this angular momentum label now refers to 
the subgroup SU(2)$_{12}$ of matrices mixing modes 1 and 2. 
Let us denote by $${\tilde D}^{(p,q)}_{(111)J_{12};(111)1}$$ the group functions obtained when working with basis states labeled
using $I_{12}$.  Some details concerning these functions, and their connection with the usual $D$ functions, can be found 
at the end of Appendix \ref{appendixsu3} and in Eqs.~(\ref{Dtildefirst}) and (\ref{Dtildesecond}).

By simple inspection we anticipate that the coefficients $A, B$ and $C$ of Eqs.~(\ref{A12})--(\ref{C12})
have an expression in terms of $\tilde D^{(p,q)}$ functions given by Eq.~(\ref{eq:23symmetry}),
provided that we replace in Eq.~(\ref{eq:23symmetry}) the usual $D^{(p,q)}$ functions defined in \cite{RSdG99}
by the corresponding $\tilde D^{(p,q)}$: 
\begin{equation}
	D^{(p,q)}_{(111)J_{23};(111)1}\to {\tilde D}^{(p,q)}_{(111)J_{12};(111)1}.  \label{substitutionrule}
\end{equation} 
This substitution rule is in fact correct:
we thus find that, when two photons are identical, the expression for the rate is ``covariant''.
The term ``covariant'' is used in the sense that the expression is
equivalent to Eq.~(\ref{eq:2indenticalphotons}) but where, in the expressions for~$A$, $B$ and $C$, 
Eq.~(\ref{substitutionrule}) is substituted and the delay $\tau$ is now interpreted as the delay between photon
3 and the simultaneous arrival of photons 1 and 2.
In general, the $\tilde D^{(p,q)}$ functions in the $I_{12}$ basis are linear combinations of 
the $D^{(p,q)}$ in the $I_{23}$ basis.
The explicit substitution of Eq.~(\ref{substitutionrule}) is easily obtained following ~\cite{RSdG99}
and is given explicitly in Appendix~\ref{appendixSU3}.

The same reasoning applies to the case that photons~1 and~3 are identical:
$(\Delta,-\Delta)$.
The expression for the coincidence rate~$\wp$ is most simply expressed now in terms of $\bar D$ functions where $I_{13}$ is a good quantum number;
again only states with $I_{13}=1$ can appear at the input.  The $\bar D$ functions for $I_{13}$ 
are again linear combinations of those where $I_{23}$ is a good quantum number.

We conclude our analysis of the case where  two or more photons are indistinguishable with the following observation:
the rate depends on four distinct $D^{\Yboxdim{4pt}\Yvcentermath1 \yng(2,1)}$ functions (in any basis) as well as one 
$D^{\Yboxdim{4pt}\Yvcentermath1 \yng(3)}$ function, so there are five functions in total.
However, we can obtain at most four rate equations.
The first three rate equations are obtained when the photon pairs $(12)$, $(13)$, and
$(23)$ are made to be indistinguishable and when the last rate equation is obtained by requiring that all photons are indistinguishable.
This last rate is proportional to the permanent alone.
 
Assuming that we have inferred the permanent from the empirical rate when all delays are~$0$,
and then we use this value in the remaining three equations,
we are still left with four distinct $D^{\Yboxdim{4pt}\Yvcentermath1 \yng(2,1)}$.
Thus, we have more unknown functions than equations.
From these considerations we see that it is not possible to completely solve the resultant coupled nonlinear quadratic equations 
and find all the immanants and the permanent when two or mode photons are identical.  
Gr\"{o}bner basis methods (as implemented, for instance, in Mathematica$^\circledR$)~\cite{Laz83}
could be used to solve for three 
$\vert D^{\Yboxdim{4pt}\Yvcentermath1 \yng(2,1)}\vert^2$ in terms of the fourth 
$\vert D^{\Yboxdim{4pt}\Yvcentermath1 \yng(2,1)}\vert^2$ and $\vert D^{\Yboxdim{4pt}\Yvcentermath1 \yng(3)}\vert^2$ 
(although the solution is not unique,
and it is not yet clear how to choose the correct one).

\subsection{All distinguishable photons}
\label{subsec:distinguishable}

We limit our discussion of this case to the case where the three photons are equally spaced in time.
This can be accomplished by setting $\bm{\Delta}=(\Delta,\Delta)$ for $\Delta$ sufficiently large compared to the pulse duration.
The coincidence rate is then
\begin{align}
	\wp\left((\Delta,\Delta);\Omega\right)
		=&C_A+\frac{1}{6}\text{e}^{-4\sigma^2\Delta^2}\Bigg(\left|D^{\Yboxdim{4pt}\Yvcentermath1 \yng(3)}_{(111)1;(111)1}\right|^2
							\nonumber\\&
			-\left(D^{\Yboxdim{4pt}\Yvcentermath1 \yng(1,1,1)}_{(111)0;(111)0}\right)^2+C_B\Bigg)
							\nonumber\\&
			+\frac{1}{2}\text{e}^{-3\sigma^2\Delta^2}\Bigg(\left|D^{\Yboxdim{4pt}\Yvcentermath1 \yng(3)}_{(111)1;(111)1}\right|^2
							\nonumber\\&
			+\left(D^{\Yboxdim{4pt}\Yvcentermath1 \yng(1,1,1)}_{(111)0;(111)0}\right)^2-2C_A\Bigg)
							\nonumber\\&
			+\frac{1}{3}\text{e}^{-\sigma^2\Delta^2}\Bigg(\left|D^{\Yboxdim{4pt}\Yvcentermath1 \yng(3)}_{(111)1;(111)1}\right|^2
							\nonumber\\&
				-\left(D^{\Yboxdim{4pt}\Yvcentermath1 \yng(1,1,1)}_{(111)0;(111)0}\right)^2-\frac{1}{2}C_B\Bigg)
\end{align}
with
\begin{equation}
	C_A=\sum_{i\neq j\neq k\neq i}|a_{ijk}|^2
\end{equation}
and
\begin{align}
	C_B=&\vert D^{{\Yboxdim{4pt}\Yvcentermath1 \yng(2,1)}}_{(111)0;(111)0}\vert^2
	-2\sqrt{3}D^{\Yboxdim{4pt}\Yvcentermath1 \yng(2,1)}_{(111)0;(111)0}D^{\Yboxdim{4pt}\Yvcentermath1 \yng(2,1)}_{(111)0;(111)1}
			\nonumber\\&
	-\vert D^{\Yboxdim{4pt}\Yvcentermath1 \yng(2,1)}_{(111)0;(111)1}\vert^2
	+\vert D^{\Yboxdim{4pt}\Yvcentermath1 \yng(2,1)}_{(111)1;(111)0}\vert^2
							\nonumber\\&
	-2\sqrt{3}D^{\Yboxdim{4pt}\Yvcentermath1 \yng(2,1)}_{(111)1;(111)0}D^{\Yboxdim{4pt}\Yvcentermath1 \yng(2,1)}_{(111)1;(111)1}
			\nonumber\\&
	-\left\vert D^{\Yboxdim{4pt}\Yvcentermath1 \yng(2,1)}_{(111)1;(111)1}\right\vert^2,
\label{eq:distinct}
\end{align}
where we note that the  $D^{\Yboxdim{4pt}\Yvcentermath1 \yng(2,1)}_{(111)J;(111)I}$ are real for all values of $\Omega$.
Equation~(\ref{eq:distinct}) appears to contribute a fourth equation in addition to the three equations from Section 
\ref{subsec:twoandone} that would allow us to solve for all four $D^{(1,1)}$ functions.
Surprisingly, Eq.~(\ref{eq:distinct}) can be written as a linear combination of the three rates with two identical time delays and hence does not contribute additional information that can be used towards a solution.

\begin{figure}
	\includegraphics[width=\columnwidth]{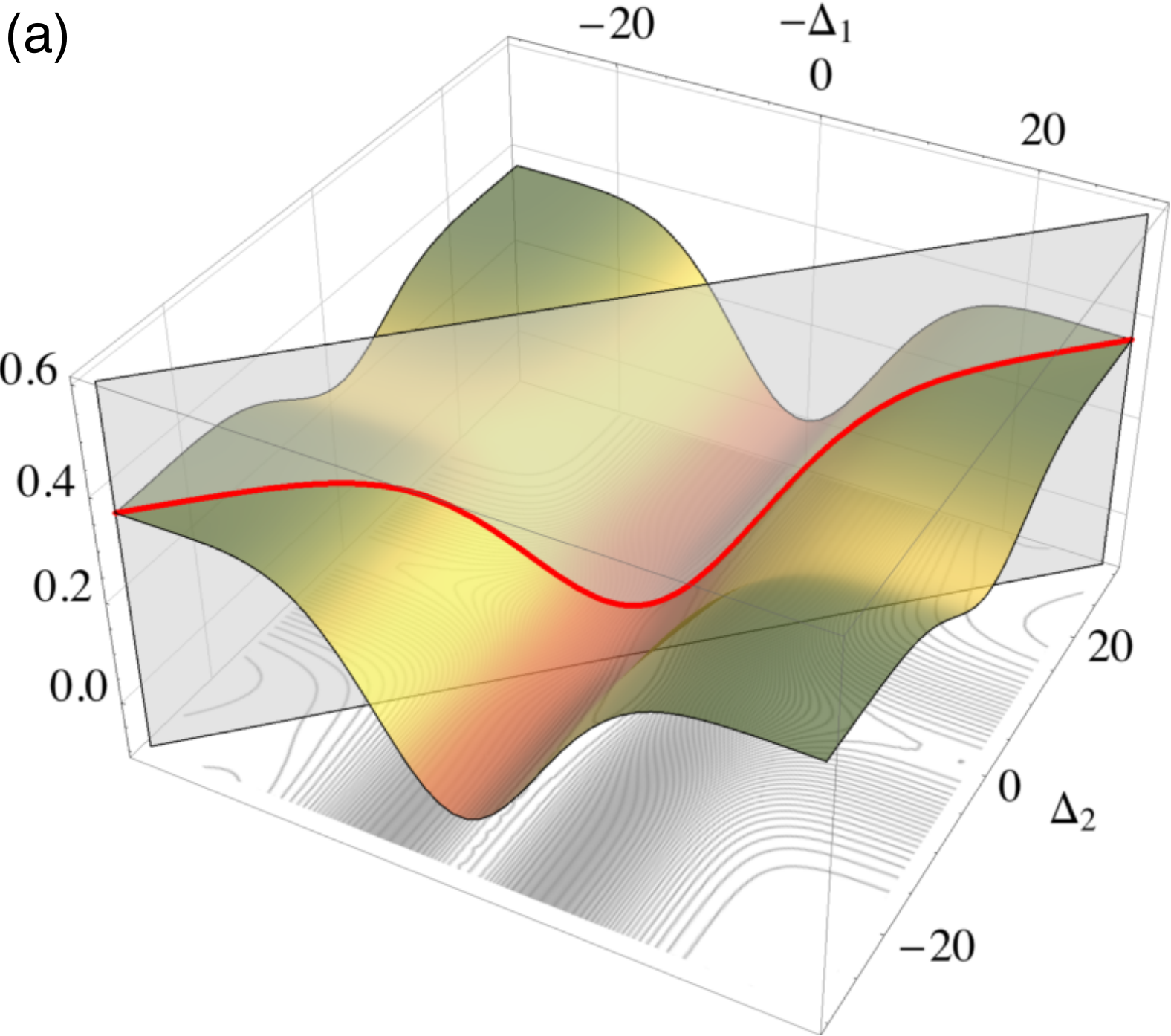}
	\includegraphics[width=\columnwidth]{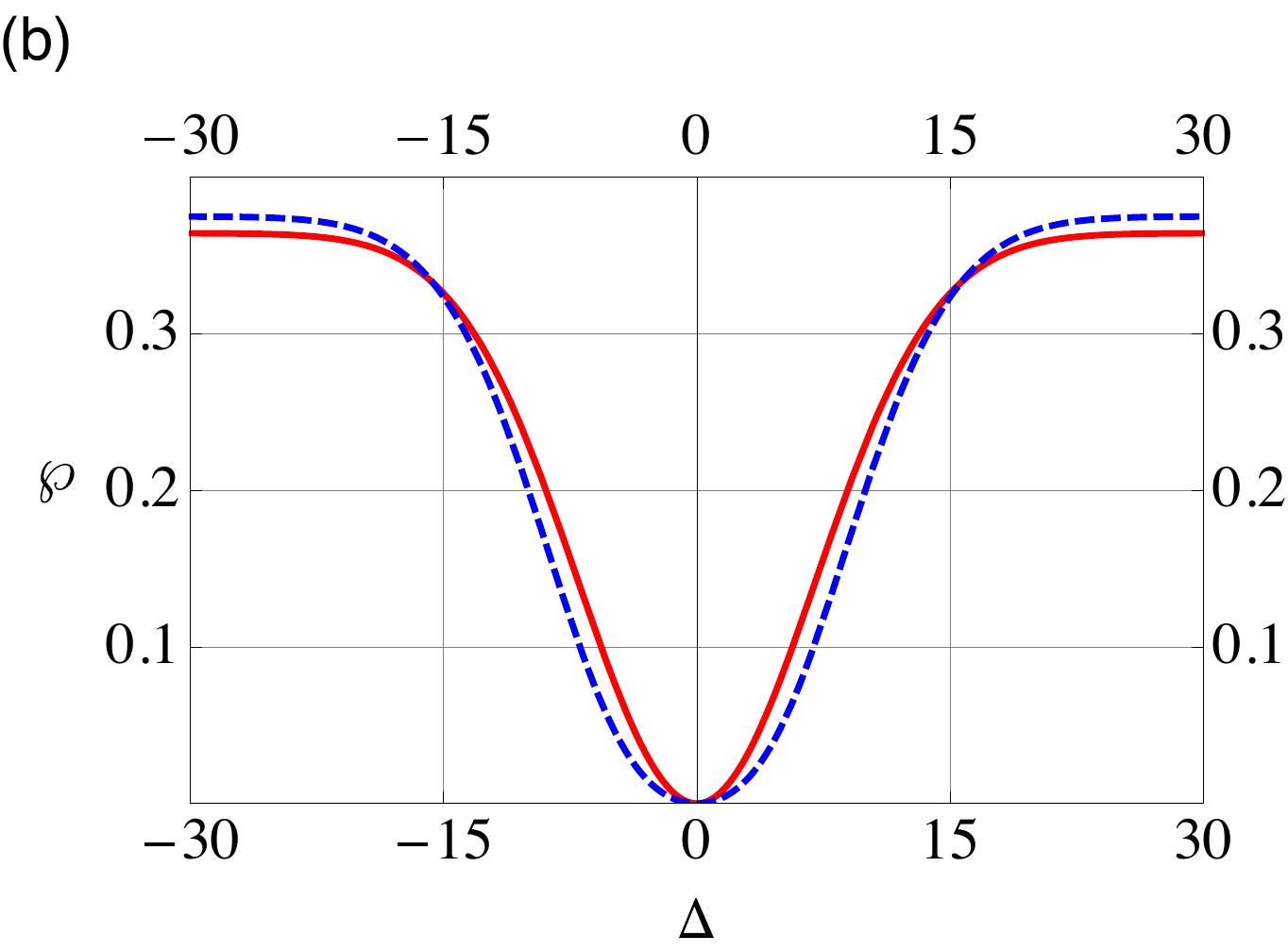}
	\caption{
    	(Color online)
Three-photon coincidence rate $\wp(\mathbf{\Delta},\Omega)$ for a three-
photon passive optical interferometer with one photon entering each input 
port shown. The rate is shown as (a) a surface plot, with the solid (red) 
line corresponding to $(\Delta,-\Delta)$ for photons 1 and 3 arriving at 
the same time, and (b) the $(\Delta,-\Delta)$ and $(\Delta,\Delta)$ lines 
as the solid (red) and dashed (blue) loci, respectively. Here~$\Omega=(\pi/3,0,\pi/5,\pi/2,\pi/3,\pi/4,0,0)$, and the single-photon spectral widths are identically $\sigma_0=0.1$.
    }
    \label{fig:nonsym}
\end{figure}

For comparison, we plot the coincidence rate for the same interferometer but with three different photon frequencies in Fig.~\ref{fig:nonsym}.
We can clearly see from Fig.~\ref{fig:nonsym}(b) that the backgrounds given at $\Delta\to\infty$ of the diagonal and antidiagonal lines are different. The diagonal is given by a Gaussian,
whereas the antidiagonal is a linear combination of Gaussians.
Of course at $\tau=0$ both the diagonal and the antidiagonal collapses to a single value and that is the modulus square of the permanent.

\section{Conclusions}

We have developed a theory and a formalism for studying three-photon coincidence rates at the output of a three-channel passive optical 
interferometer.
The input is three photons, one of which enters each of the three input ports of the interferometer.
The photons are in pulse modes in order to ensure that controllable delays can be applied to each photon independently.
Other than the delay times, the photons are treated as identical in every way.
The three-photon coincidence rate is calculated by using integrals over frequency modes and exploiting permutation groups,
SU(3) Lie group theory,
representation theory,
and the theory of immanants, which includes determinants and permanents of matrices as special cases.

The analysis we present here builds on our earlier brief study of non-simultaneous identical photons and their coincidences in 
passive three-channel optical interferometry,
but here we elaborate on the many technical aspects and study asymptotic behavior, which helps us to characterize and understand
the photon-coincidence-rate landscape.
Furthermore, we employ here a distinct description of the photon counters:
in contrast to our earlier work, which employed an idealistic dualism between source photons and photon detection,
here we discuss the coincidence-rate landscape in terms of the measurement operator corresponding to currently used detectors.

A key contribution of our work is as a generalization of the Hong-Ou-Mandel dip,
which is one of the most important demonstrations and tools used in quantum optics.
The Hong-Ou-Mandel dip phenomenon hinges on the observation that identical photons entering two ports of a balanced beam splitter
yield an output corresponding to a superposition of both photons exiting the two ports together in tandem.
Experimentally the dip is observed by varying the relative delay time between the arrival of the two photons,
thus controlling their mutual degree of 
distinguishability from indistinguishable where the photon arrivals are simultaneously to completely distinguishable when the photon arrivals are 
separated by more than the duration of the photon pulses.

We have generalized to controllable distinguishability of the three photons entering a general three-channel passive linear optical interferometer.
Although this controllability is desirable for practical reasons,
the mathematics used to describe this three-photon generalization of the Hong-Ou-Mandel dip is nontrivial and beautiful in
its application of group theory.

Our work shows the path forward to considering more photons entering interferometers with at least as many channels as photons,
which is the case of interest for BosonSampling.
Whereas the BosonSampling problem is framed in the context of simultaneous photon arrival times,
thereby leading to matrix permanents in the sampling computations,
our work opens BosonSampling to nonsimultaneity of photons, hence the role of immanants in the sampling of photon coincidence rates.
The case of three photons in three modes is the simplest situation where the theory requires immanants beyond the permanent and the
determinant.

In summary, our work generalizes the Hong-Ou-Mandel dip to the three-photon, three-channel case
and points the way forward to analyze further multi-photon, multichannel generalizations.
Our work is important for characterizing and understanding the consequent photon-coincidence landscapes.
In addition, our use of group theory to study controllable delays in photon arrival times shows how the BosonSampling device
can yield rates that depend on matrix immanants, which generalizes the matrix permanent analysis in the original BosonSampling studies.

\acknowledgements
We thank A.~Branczyk, I.~Dhand, M.~Tichy, and M.~Tillman for helpful 
discussions. H.d.G and I.P.P. acknowledge support from NSERC and Lakehead 
University. S.H.T. acknowledges that this material is based on research supported 
in part by the Singapore National Research Foundation under NRF Award No. 
\ NRF-NRFF2013-01. B.C.S. is supported by NSERC, USARO, and AITF
and acknowledges hospitality and financial support from Macquarie 
University in Sydney and from the Raman Research Institute in Bangalore, where some of this research was performed.

\appendix

\section{Essentials concerning $\mathfrak{su}$(3) and SU(3)}\label{appendixsu3}

The Lie algebra $\mathfrak{su}$(3) is spanned by six ladder operators 
\begin{equation}
	\hat C_{12},  \hat C_{13}, \hat C_{23}, \hat C_{21},  \hat C_{31}, \hat C_{32}
\end{equation}
and two commuting `weight' operators expressed here as
\begin{equation}
	\hat h_1:=\hat C_{11}-\hat C_{22},\;
	\hat h_2:=\hat C_{22}-\hat C_{33}.
\end{equation}
The operators~$\hat C_{ij}$ satisfy the commutation relations
\begin{equation}
	\left[\hat C_{ij},\hat C_{k\ell}\right]=\hat C_{i\ell}\delta_{jk}-\hat C_{kj}\delta_{\ell i}.
\end{equation}

Within the context of our work, it is convenient to realize these operators in terms of photon creation and destruction operators
as 
\begin{equation}
\label{eq:abstractCij}
\hat C_{ij}=\hat a^\dagger_i (\omega_1) \hat a_j (\omega_1) + \hat a^\dagger_i (\omega_2) \hat a_j (\omega_2)
+\hat a^\dagger_i (\omega_3) \hat a_j (\omega_3).
\end{equation}
Note that $\hat C_{ij}$ is invariant (unchanged) by permutation of the frequencies.  Thus, if a state is constructed to 
have specific symmetries under permutation of the frequencies, 
the action of $\hat C_{ij}$ maps this state to another \emph{having the same 
specific symmetries under permutation of the frequencies.}

Of fundamental importance in representations of $\mathfrak{su}$(3) is the so--called highest-weight state.  This is a state
annihilated by all the ``raising operators'': $\hat C_{12}, \hat C_{13}$, and $\hat C_{23}$.  
For instance, the states
\begin{align}
\label{eq:twohighestweights-1}
	\left\vert \hbox{\small\Yvcentermath1$\young(12,3)$}\right\rangle 
		=&\frac{1}{\sqrt{2}} \Big(\hat a_1^\dagger(\omega_1) \hat a_2^\dagger(\omega_3)
				\nonumber\\&
			- \hat a_2^\dagger(\omega_1) 
\hat a_1^\dagger(\omega_3) \Big)
\hat a_1^\dagger(\omega_2)\vert 0\rangle
\end{align}
and
\begin{align}
\label{eq:twohighestweights-2}
	\left\vert \hbox{\small\Yvcentermath1$\young(13,2)$}\right\rangle
		=&\frac{1}{\sqrt{2}} \Big(\hat a_1^\dagger(\omega_1) \hat a_2^\dagger(\omega_2) 
				\nonumber\\&
	-\hat a_2^\dagger(\omega_1) \hat a_1^\dagger(\omega_2) \Big)
\hat a_1^\dagger(\omega_3)\vert 0\rangle
\end{align}
are both highest-weight states (under the action of the raising operators given in Eq.~(\ref{eq:abstractCij}).)

The weight of a state is the vector $(p,q)$ of eigenvalues of the operators $\hat h_1$ and 
$\hat h_2$.  In terms of occupation number $(n_1,n_2,n_3)$, the weight of a state is therefore
simply  $(n_1-n_2,n_2-n_3)$ and is frequency independent.  

The two states of Eqs.~(\ref{eq:twohighestweights-1}) and~(\ref{eq:twohighestweights-2}) both have weight $(1,1)$.
Because all the states of a representation can be obtained by repeatedly acting on the highest-weight state
using the lowering operators 
$\hat C_{21},\hat C_{31}$, and $\hat C_{32}$, the weight of the highest weight state is used to label 
states in the whole
representation.   

For finite-dimensional unitary representations of $\mathfrak{su}$(3), 
one can always choose the components $(p,q)$ of the highest-weight to be non-negative
integers.  
The dimensionality of the representation $(p,q)$ is
\begin{equation}
	(p+1)(q+1)(p+q+2)/2,
\end{equation}
so that dim$[(1,1)]=8$.

The two states of Eqs.~(\ref{eq:twohighestweights-1}) and~(\ref{eq:twohighestweights-2}) are not orthogonal; however, 
since a linear combination of those states is also a highest-weight state,
it is possible to orthonormalize them using the
usual Gram-Schmidt method.  For instance,
\begin{align}
\ket{1}&=\left\vert \hbox{\small\Yvcentermath1$\young(12,3)$}\right\rangle \, , \label{rep1}\\
\ket{2}&=\frac{1}{\sqrt{3}}\left\vert \hbox{\small\Yvcentermath1$\young(12,3)$}\right\rangle
-\frac{2}{\sqrt{3}}\left\vert \hbox{\small\Yvcentermath1$\young(13,2)$}\right\rangle.
\end{align}
These can serve as distinct highest-weight states for the two distinct copies of the irrep $(1,1)$ or 
${\Yboxdim{6pt}\Yvcentermath1 \yng(2,1)}$ that occur in the
decomposition of our Hilbert state.  
Obviously the choice of $\ket{1}$ and $\ket{2}$ as highest-weight states with weight (1,1) is not unique, but 
all other highest-weight states with weight (1,1) can be written
as a linear combination of $\ket{1}$ and $\ket{2}$; if not, there would be a third copy of (1,1) in the
Hilbert space.  

The matrix representations of elements of $\mathfrak{su}$(3) obtained using either highest-weight
state is equivalent ; i.e., they differ by at most a common unitary change of basis.   Nevertheless, any state
obtained by lowering operators acting on $\ket{1}$ are always orthogonal to states obtained by lowering operators
acting on $\ket{2}$.

Now consider states of the form
\begin{align}
\label{ijsingletstate}
	\ket{(1,1)111;0}_{ijk}
		=& \frac{1}{\sqrt{2}}\Big(\hat a_2^\dagger(\omega_i)\hat a_3^\dagger(\omega_j)
					\nonumber\\&
			-\hat a_2^\dagger(\omega_j)\hat a_3^\dagger(\omega_i)\Big)\hat a_1^\dagger(\omega_k)\vert 0\rangle,
\end{align}
for~$i\ne j\ne k\neq i$.

The triple $111$ indicates that they are constructed as superpositions of states
with one quantum in each mode; the weight of these states is $(0,0)$.  
They are obviously antisymmetric under permutation of modes
2 and 3, and under permutation of frequencies
$\omega_i$ and $\omega_j$.  They are also annihilated by the operators $\hat C_{23}$ and $\hat C_{32}$;
they are eigenstates of $\hat h_2$ with eigenvalue 0.  If we observe
that the operators $\{\hat C_{23},\hat C_{32},\frac{1}{2}\hat h_2\}$ have the same commutation relations as the angular 
momentum operators, we conclude that $\ket{(1,1)111;0}_{ijk}$ are in fact states of angular momentum 
$I_{23}=0$ (i.e., singlet) states.  This is the interpretation of the last index $0$ in the states.

States with weight $(0,0)$ and $I=0$ in both 
${\Yboxdim{6pt}\Yvcentermath1 \yng(2,1)}$ representations are linear combinations of the 
$\ket{(1,1)111;0}_{ijk}$ states.
For instance, the state
\begin{align}
\label{I230state}
	\left|(1,1)111;0\right\rangle_1
		=&-\frac{\ket{(1,1)111;0}_{231}}{\sqrt{6}}
				\nonumber\\&
			-\frac{\sqrt{2}\ket{(1,1)111;0}_{132}}{\sqrt{3}}
				\nonumber\\&
			+\frac{\ket{(1,1)111;0}_{213}}{\sqrt{6}}
\end{align}
is in the representation having $\ket{1}$ of Eq.~(\ref{rep1}) as the highest weight.
As a linear combination of 
states antisymmetric under exchange of modes 2 and 3, $\left|(1,1)111;0\right\rangle_1$ is itself antisymmetric under such exchange.

On the other hand, states of the form
\begin{align}
\label{ijtripletstate}
	\ket{(1,1)111;1}_{ijk}
		=&\frac{1}{\sqrt{2}}\Big(\hat a_2^\dagger(\omega_i)\hat a_3^\dagger(\omega_j)
				\nonumber\\&
			+\hat a_2^\dagger(\omega_j)\hat a_3^\dagger(\omega_i)\Big)\hat a_1^\dagger(\omega_k)\vert 0\rangle,
\end{align}
where $i\ne j\ne k\neq i$ can be shown to have angular momentum $I_{23}=1$.  They are symmetric under 
permutation of modes 2 and 3 and under permutation of frequencies
$\omega_i$ and $\omega_j$.   

States with weight $(0,0)$ and $I=1$ in both 
${\Yboxdim{6pt}\Yvcentermath1 \yng(2,1)}$ representations are linear combinations of the 
$\ket{(1,1)111;1}_{ijk}$ states.
For instance, the state
\begin{equation}
\vert (1,1)111;1\rangle_1= \frac{\ket{(1,1)111;1}_{231}}{\sqrt{2}}-\frac{\ket{(1,1)111;1}_{213}}{\sqrt{2}} \label{I231state}
\end{equation}
is in the representation having $\ket{1}$ of Eq.~(\ref{rep1}) as the highest weight.  As it is constructed from states explicitly
symmetric under exchange of modes 2 and 3, $\vert (1,1)111;1\rangle_1$ is itself symmetric under this permutation of modes.

Hence we see a feature of the irrep 
${\Yboxdim{6pt}\Yvcentermath1 \yng(2,1)}$ of $\mathfrak{su}$(3) that
does not occur in angular momentum theory: it is possible
to have distinct states such as $\vert (1,1)111;0\rangle_1$ and $\vert (1,1)111;1\rangle_1$,
 with the same weight ; i.e. the weight is not enough to uniquely identify the state.  
 [This multiplicity of weight \emph{never} occurs in $\mathfrak{su}(2)$, where the integral weight $2m$ is enough to
 completely identify the state in the irrep.]
In addition to the weight,
one must in general supply an additional index, $I_{23}$.  In $\mathfrak{su}$(3) representations of type $(p,0)$ or $(0,q)$ this extra label 
is not necessary and often not indicated.

The states $\vert(1,1)111;0\rangle_{ijk}$ and $\ket{(1,1)111;1}_{ijk}$ of Eqs.~(\ref{ijsingletstate}) and
(\ref{ijtripletstate}) 
are not the only possible states that can be used to construct zero-weight states with desirable permutation symmetries:
labeling states with the weight using $I_{23}$ is not the only possible choice.
We can consider, for instance,
\begin{align}
&\widetilde{\left|(1,1)111;0\right\rangle}_{ijk}
	\nonumber \\&
		= \left(\hat a_1^\dagger(\omega_i)\hat a_2^\dagger(\omega_j)- \hat a_2^\dagger(\omega_j)\hat a_1^\dagger(\omega_i)
\right)\hat a_3^\dagger(\omega_k)\vert 0\rangle
\label{barijsingletstate}
\end{align}
and
\begin{align}
&\widetilde{\ket{(1,1)111;1}}_{ijk}\nonumber \\
&\quad = 
\left(\hat a_1^\dagger(\omega_i)\hat a_2^\dagger(\omega_j)+\hat a_2^\dagger(\omega_j)\hat a_1^\dagger(\omega_i)
\right)\hat a_3^\dagger(\omega_k)\vert 0\rangle.\label{barijtripletstate} 
\end{align}
These states are now obviously states of angular momentum $I_{12}=0$ and $I_{12}=1$ respectively, where the
angular momentum algebra $\mathfrak{su}(2)_{12}$ is spanned by 
$\{\hat C_{12},\hat C_{12},\frac{1}{2}\hat h_1\}$.

The states~(\ref{barijsingletstate}) and (\ref{barijtripletstate}) can be 
used to construct an alternative basis for the weight-$0$ subspace of the 
irrep with highest weight state $\ket{1}$. In other words states~$
\widetilde{\ket{(1,1)111;0}}_1$ and~$\widetilde{\ket{(1,1)111;1}}_1$
can be defined such that they carry the angular momentum labels $I_{12}=0$ 
and $1$ respectively, defined in terms of $\mathfrak{su}(2)_{12}$.
These states are appropriate linear combinations of 
$\widetilde{\ket{(1,1)111;0}}_{ijk}$ or $\widetilde{\ket{(1,1)111;1}}
_{ijk}$ states and so antisymmetric (respectively symmetric) under 
exchange of modes~1 and~2.
The group functions defined in terms of basis states like
$$\widetilde{\ket{(1,1)111;0}}_1\text{ and }\widetilde{\ket{(1,1)111;1}}
_1$$ with $I_{12}$ labeling
the angular momentum properties of the states are denoted $\tilde 
D^{(p,q)}_{(111)J_{12};(111)I_{12}}$.

Using $\mathfrak{su}(2)_{12}$ to label states represents a change of basis from a previous labeling scheme~\cite{RSdG99},
where
$\mathfrak{su}(2)_{23}$ is used.
Thus, the states in  $\mathfrak{su}(2)_{12}$ are linear combinations of those in $\mathfrak{su}(2)_{23}$,
so that $$\tilde D^{(p,q)}_{(111)J_{12};(111)I_{12}}$$ are linear combinations of the $$D^{(p,q)}_{(111)J;(111)I}$$ states used previous~\cite{RSdG99}
and in Appendix \ref{appendixSU3}.
Some explicit examples of transformations, required for our analysis, are given in 
Eqs.~(\ref{Dtildefirst}) and (\ref{Dtildesecond}).

Finally, we note that it is also possible, following exactly the same procedure as above,
 to use the subalgebra $\mathfrak{su}(2)_{13}$ to label states.
 This procedure corresponds to just another change of basis,
and the resulting $D$ functions are denoted $\bar D$.

\section{Explicit expression of some SU(3) D-functions}\label{appendixSU3}

Some functions of the $D^{(p,q)}_{(111)J;(111)I}$ type useful in constructing permanents and immanents are listed in Table III.

\begin{widetext}
\begin{table}[t]
\parbox{\textwidth}{\caption{Functions of the $D^{(p,q)}_{(111)J;(111)I}$ type useful in constructing permanents and immanants.}}
\label{table:Dpq}
{\renewcommand{\arraystretch}{1.75}
\begin{tabularx}{\textwidth}{ c @{\hspace{2cm}} c @{\hspace{2cm}} c @{\hspace{2cm}} c }
\hline \hline
$(p,q)$ &Diagram &$ (J,I)$ & $D^{(p,q)}_{(111)J;(111)I}(\Omega)$ \\
\hline \hline
(3,0) &{\small \yng(3)} & (1,1) &
$\cos \beta_1 \cos \beta_2 \cos \beta_3$ \\
&&&$-\frac{1}{4}\sin\beta_1\cos\frac{\beta_2}{2}\sin\beta_3\left(3\text{e}^{i (\alpha_1-\alpha_3)} \cos\beta_2
-\text{e}^{ i (\alpha_1-\alpha_3)}+
2\text{e}^{- i (\alpha_1-\alpha_3)}\right)
$\\
(1,1)&{\small \Yvcentermath1  $\yng(2,1)$} & (1,1) &
$
\frac{1}{4}  \left( \cos \beta_1 \cos \beta_3(\cos \beta_2+3) 
-4 \cos(\alpha_1-\alpha_3)\, \sin \beta_1 \cos\frac{\beta_2}{2} 
\sin \beta_3\right)
$\\
&&(1,0)&
$-\frac{1}{2} \sqrt{3} \cos \beta_1\sin ^2 \frac{\beta_2}{2} $\\
&&(0,1)&
$-\frac{1}{2} \sqrt{3} \sin ^2 \frac{\beta_2}{2} \cos \beta_3$\\
&&(0,0)&
$\frac{1}{4} (3 \cos \beta_2+1)$ \\
\hline \hline
\end{tabularx}
}
\end{table}
\end{widetext}

The functions $$\tilde D^{\Yboxdim{4pt}\Yvcentermath1 \yng(2,1)}_{(111)J_{12};(111)I_{12}},$$
with $I_{12}$ a good quantum number,
are related by a linear transformation to
the functions $$D^{\Yboxdim{4pt}\Yvcentermath1 \yng(2,1)}_{(111)J_{23};(111)I_{23}},$$
with~$I_{23}$ a good quantum number.
Explicitly, using the notation of ~\cite{RSdG99}, the states in the $I_{12}$ basis are defined by
\begin{equation}
\vert (1,1)111;I_{12}\rangle := \left[\psi_{\nu_3}(3)\otimes\left[\psi_{\nu_1}(1)\otimes\psi_{\nu_2}(2)\right]^{I_{12}}\right]^{1/2}_{1/2}
\end{equation} 
and can be expanded in terms of the $I_{23}$ states so that
\begin{widetext}
\begin{align}
	\vert (1,1)111;I_{12}=0\rangle =&
-\frac{1}{2}\vert (1,1)111;I_{23}=0\rangle+\frac{\sqrt{3}}{2}\vert (1,1)111;I_{23}=1\rangle\, , \\
	\vert (1,1)111;I_{12}=1\rangle =& 
-\frac{\sqrt{3}}{2}\vert (1,1)111;I_{23}=0\rangle -\frac{1}{2}\vert (1,1)111;I_{23}=1\rangle.
\end{align}
Therefore,
\begin{align}
	\tilde D^{\Yboxdim{4pt}\Yvcentermath1 \yng(2,1)}_{(111)0;(111)1}
		=&\frac{\sqrt{3}}{4}D^{\Yboxdim{4pt}\Yvcentermath1 \yng(2,1)}_{(111)0;(111)0}-\frac{\sqrt{3}}{4}
		D^{\Yboxdim{4pt}\Yvcentermath1 \yng(2,1)}_{(111)1;(111)1}
		-\frac{3}{4} D^{\Yboxdim{4pt}\Yvcentermath1 \yng(2,1)}_{(111)1;(111)0} +\frac{1}{4} 
		D^{\Yboxdim{4pt}\Yvcentermath1 \yng(2,1)}_{(111)0;(111)1} \label{Dtildefirst} , \\		
		\tilde D^{\Yboxdim{4pt}\Yvcentermath1 \yng(2,1)}_{(111)1;(111)1}
		=&\frac{3}{4}D^{\Yboxdim{4pt}\Yvcentermath1 \yng(2,1)}_{(111)0;(111)0}+\frac{\sqrt{3}}{4} 
		D^{\Yboxdim{4pt}\Yvcentermath1 \yng(2,1)}_{(111)0;(111)1}
		+\frac{ \sqrt{3}}{4} D^{\Yboxdim{4pt}\Yvcentermath1 \yng(2,1)}_{(111)1;(111)0}
		+\frac{1}{4}D^{\Yboxdim{4pt}\Yvcentermath1 \yng(2,1)}_{(111)1;(111)1}.
\label{Dtildesecond}
\end{align}
\end{widetext}
and
\begin{equation}
\tilde D^{\Yboxdim{4pt}\Yvcentermath1 \yng(3)}_{(111)1;(111)1}=  D^{\Yboxdim{4pt}\Yvcentermath1 \yng(3)}_{(111)1;(111)1}.
\end{equation}
As discussed in the text and Appendix~\ref{appendixsu3}, these functions are useful when analyzing the symmetry properties of states under
permutations of modes 1 and 2.

\section{Explicit expression of rates}\label{rateD}
In order to obtain a general expression for the rate, we need first to expand the square of the modulus of Eq.~(\ref{fullrate3photoncase}) and
then integrate.
The resultant expression thus contains sums of products of the type $$(a_{ijk})^* (a_{i'j'k'}) M_{ijk,i'j'k'},$$
where $M_{ijk,i'j'k'}$ is a factor obtained by integration of the frequencies in the term containing $(a_{ijk})^* (a_{i'j'k'})$.  
These $M_{ijk,i'j'k'}$ factors can be collected in a matrix
with column labeled by $a_{ijk}$ and rows labeled by $(a_{ijk})^* $.

Upon integration,
Eq.~(\ref{fullrate3photoncase}) for the three-photon coincidence rate
is written explicitly in terms of the time-of-arrival vector $\bm\tau$ of the photons in mode 1, 2, and 3 respectively.
For $\bm{\tau}$ and $\bm\Delta$ related by expression~(\ref{eq:tauDelta}),
this rate is given by
\begin{align}
	\wp(\bm{\Delta};\Omega)
		=& \bm{a}(\Omega)^\dag M_{\text{rate}}(\bm{\tau}) \bm{a}(\Omega).
\end{align}
Here $M_{\text{rate}}$ is the $6\times6$ symmetric matrix
\begin{widetext}
\begin{align}
	M_{\text{rate}}=\begin{pmatrix}
		1 & \text{e}^{-\sigma^2(\tau_2-\tau_3)^2} & \text{e}^{-\sigma^2(\tau_1-\tau_2)^2}
			&\text{e}^{-\sigma^2\tau_C^2} & \text{e}^{-\sigma^2\tau_C^2}&\text{e}^{-\sigma^2(\tau_1-\tau_3)^2}\\
		\text{e}^{-\sigma^2(\tau_2-\tau_3)^2} & 1 & \text{e}^{-\sigma^2\tau_C^2}
			&\text{e}^{-\sigma^2(\tau_1-\tau_3)^2} & \text{e}^{-\sigma^2(\tau_1-\tau_2)^2} 
			& \text{e}^{-\sigma^2\tau_C^2}\\
		\text{e}^{-\sigma^2(\tau_1-\tau_2)^2} &\text{e}^{-\sigma^2\tau_C^2} & 1
			& \text{e}^{-\sigma^2(\tau_2-\tau_3)^2} & \text{e}^{-\sigma^2(\tau_1-\tau_3)^2} 
			& \text{e}^{-\sigma^2\tau_C^2}\\
		\text{e}^{-\sigma^2\tau_C^2} &\text{e}^{-\sigma^2(\tau_1-\tau_3)^2}
			&\text{e}^{-\sigma^2(\tau_2-\tau_3)^2} & 1 & \text{e}^{-\sigma^2\tau_C^2} 
			&\text{e}^{-\sigma^2(\tau_1^2-\tau_2)^2}\\
		\text{e}^{-\sigma^2\tau_C^2} &\text{e}^{-\sigma^2(\tau_1-\tau_2)^2}
			&\text{e}^{-\sigma^2(\tau_1-\tau_3)^2}
			& \text{e}^{-\sigma^2\tau_C^2} & 1 
			& \text{e}^{-\sigma^2(\tau_2-\tau_3)^2}\\
		\text{e}^{-\sigma^2(\tau_1-\tau_3)^2} &\text{e}^{-\sigma^2\tau_C^2}
			&\text{e}^{-\sigma^2\tau_C^2} & \text{e}^{-\sigma^2(\tau_1-\tau_2)^2} 
			&\text{e}^{-\sigma^2(\tau_2-\tau_3)^2} & 1
\end{pmatrix}
\end{align}
\end{widetext}
with
\begin{equation}
\label{eq:tauC}
	\tau_C=\sqrt{|\bm{\tau}|^2-\tau_2\tau_3-\tau_1\tau_2-\tau_1\tau_3}
\end{equation}
and
\begin{align}
\label{Eq:D}
\bm{a}(\Omega)
	=\begin{pmatrix}
		a_{123}(\Omega)\\
		a_{132}(\Omega)\\
		a_{213}(\Omega)\\
		a_{231}(\Omega)\\
		a_{312}(\Omega)\\
		a_{321}(\Omega)
	\end{pmatrix}.
\end{align}
for which $\{a_{ijk}\}$ is defined in Eq.~(\ref{eq:aijk}).

\bibliographystyle{apsrev4-1.bst}

\bibliography{arxiv_v2_submission}
\end{document}